\documentclass[times, 10pt,twocolumn]{article}
\usepackage{latex8}
\usepackage{times}

\usepackage[normalem]{ulem}
\usepackage{epsfig,endnotes}
\usepackage{graphicx}
\usepackage{epstopdf}
\usepackage{setspace} 
\usepackage{subfig}
\usepackage{enumitem}
\usepackage[boxed]{algorithm2e}

\begin{document}

\newcommand{\captions}[1]{\caption{\setstretch{0.1}{#1}}}
\newcommand{\footnotes}[1]{\footnote{\setstretch{0.1}{#1}}}

\title{
Sprinkler: Maximizing Resource Utilization in Many-Chip Solid State Disks}

\author{
{\rm Myoungsoo Jung$^{1}$ \ and Mahmut T. Kandemir$^{2}$}\\ 
       $^{1}$ Computer Architecture and Memory Systems Laboratory, Yonsei University\footnotemark,    \\ 
       $^{2}$ Department of EECS, The Pennsylvania State University\\
       {\small mj@camelab.org, kandemir@cse.psu.edu}
}

\date{}
\maketitle

\thispagestyle{empty}

\begin{abstract}
Resource utilization is one of the emerging problems in many-chip SSDs. In this paper, we propose Sprinkler, a novel device-level SSD controller, which targets  maximizing resource utilization and achieving high performance without additional NAND flash chips. 
Specifically, Sprinkler relaxes parallelism dependency by scheduling I/O requests based on internal resource layout rather than the order imposed by the device-level queue. 
In addition, Sprinkler improves flash-level parallelism and reduces the number of transactions (i.e., improves transactional-locality) by over-committing flash memory requests to specific resources. Our extensive experimental evaluation using a cycle-accurate large-scale SSD simulation framework shows that a many-chip SSD equipped with our Sprinkler provides at least 56.6\% shorter latency and 1.8 \textasciitilde 2.2 times better throughput than the state-of-the-art SSD controllers. Further, it improves overall resource utilization by 68.8\% under different I/O request patterns and provides, on average, 80.2\% more flash-level parallelism by reducing half of the flash memory requests at runtime. 
\end{abstract}

\footnotetext[1]{This paper is published at 20th IEEE International Symposium On High Performance Computer Architecture and mostly done when he was at University of Texas at Dallas and Pennsylvania State University. This material includes new data (regarding garbage collection) and is presented to ensure timely dissemination of scholarly and technical work.}

\section{Introduction}
Flash-based memory cards and embedded SSDs have become the dominant storage technology in mobile devices, and large-scale SSDs are being rapidly deployed in laptops and workstations. Further, enterprise, data-intensive and high performance computing have begun to employ SSDs through high speed interfaces like PCI Express -- data rate is 16GB/sec --, in an attempt to avoid conventional storage interface overheads, and exploit the advantages brought by NAND flash \cite{spec:nvme, spec:pcie, hpc:iointensive, moneta}. 

However, since individual NAND flash device bandwidths are still around 40$\sim$400MB/sec, modern SSDs are undergoing severe architectural changes. Specifically, hundreds to thousands of NAND flash memory chips are being interconnected to form a single storage, and multiple I/O channels and cores are being integrated with these NAND flash chips. In parallel, new NAND flash technologies are being developed to extract the maximum amount of data access parallelism. A single flash chip consists of multiple dies, each of which accommodating multiple planes. Thanks to this many-chip architecture, SSDs can easily scale up and improve their performance by introducing more internal resources. As expected, performance characteristics of modern SSDs vary based on how well these hundreds or thousands of flash dies and planes are utilized. 
In order to efficiently manage these resources, different architectural approaches have been explored \cite{hpc:gordon, chen}. From a system viewpoint, techniques such as ganging \cite{arch:dtssd} and superblocking \cite{dirik} split an I/O request into multiple flash memory requests, and scatter them across different internal SSD resources. In comparison, at a flash level, \cite{arch:dtssd} and \cite{dirik} interleave incoming requests across multiple dies and planes. Various page allocation schemes \cite{pas, shin, hu} have also been investigated, which determine a physical data layout that can take advantage of internal parallelism \cite{darksecret}. All these prior proposals parallelize data accesses across abundant internal SSD resources based on incoming I/O requests, and can therefore potentially improve SSD performance. 


\begin{figure}
\centering
\def\subfigcapskip{0pt}
\subfloat[Performance Stagnation]{\label{fig:small-read-perf}\rotatebox{0}{\includegraphics[width=0.7\linewidth]{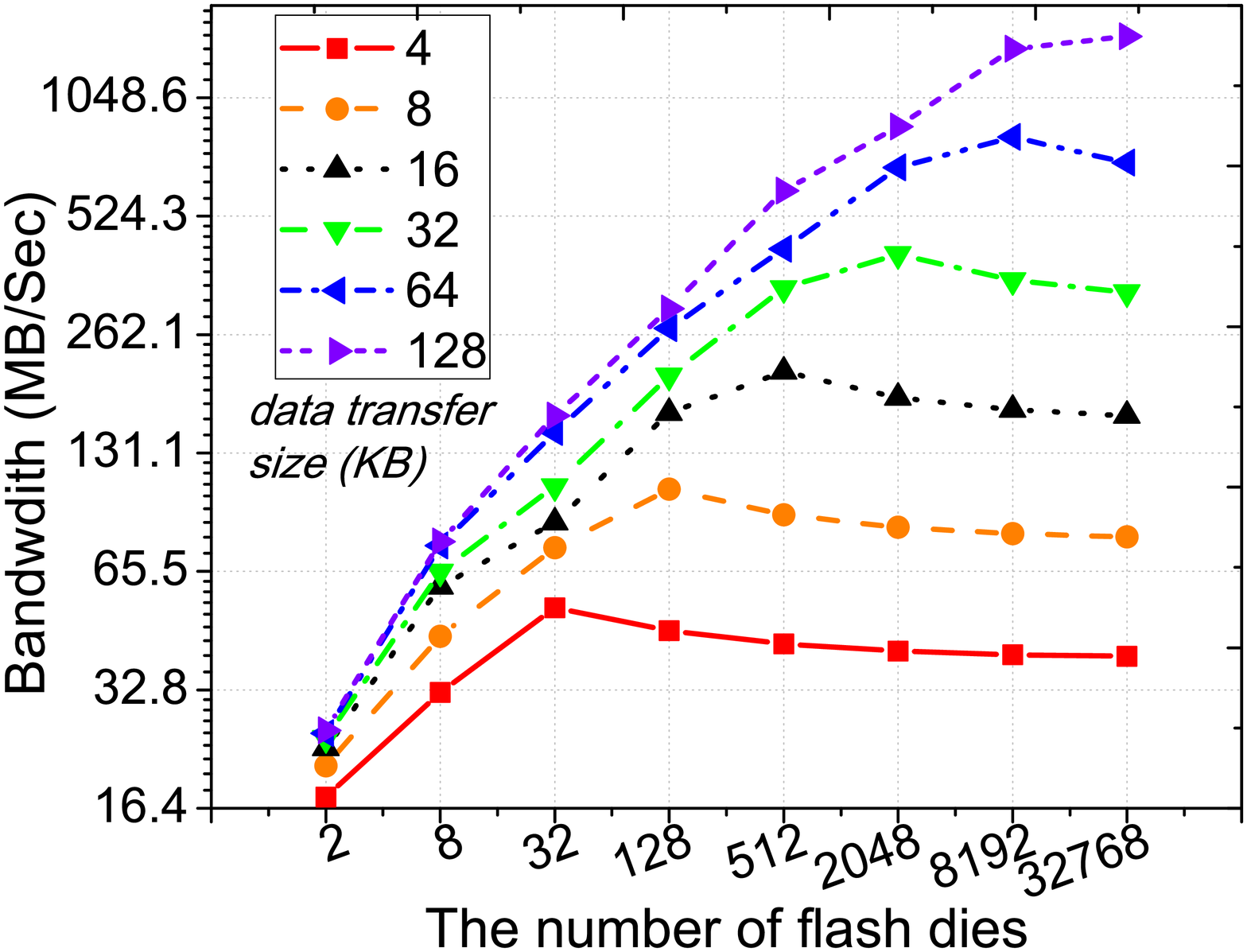}}}
\vspace{-10pt}

\subfloat[Utilization and Idleness]{\label{fig:small-read-utilidle}\rotatebox{0}{\includegraphics[width=0.7\linewidth]{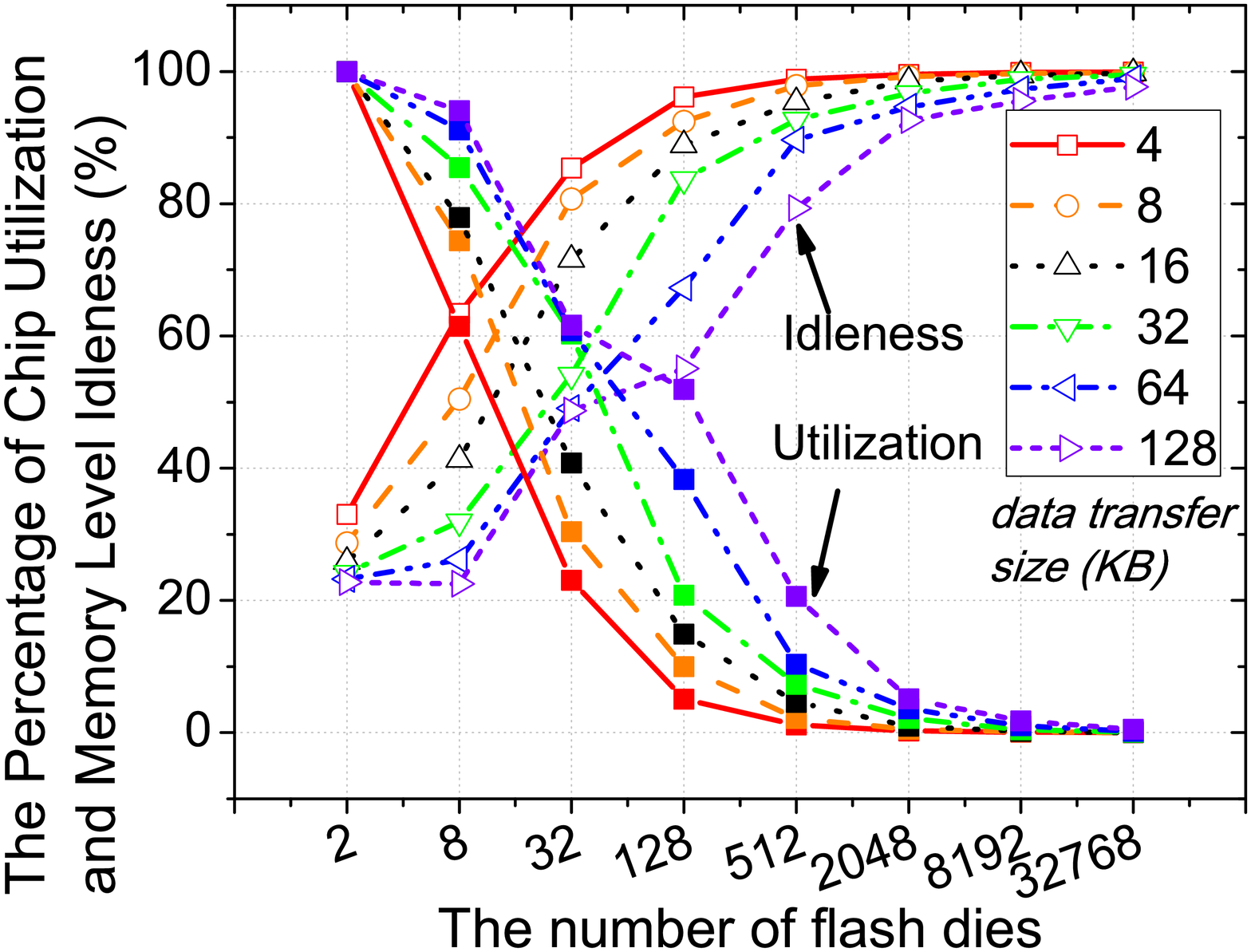}}}
\vspace{-5pt}
\captions{Many-chip SSD performance, chip utilization, and memory-level idleness sensitivity to varying number of flash chips and data transfer sizes. Each curve corresponds to a different data transfer size.}
\label{fig:many-chip-motive}
\vspace{-15pt}
\end{figure}

However, we found that, unlike common expectation, the performance of many-chip SSDs are unfortunately not significantly improved as the amount of internal resources increases, which means that employing more and more flash chips is not a promising solution. In fact, our cycle-level simulation data reveal that the read bandwidth of a state-of-the-art many-chip SSD stagnates (Figure \ref{fig:small-read-perf}), the internal resource utilization goes sharply down, and the flash memory-level idleness keeps growing (Figure \ref{fig:small-read-utilidle}), as we increase the number of dies from two to thirty thousands. 

We believe that there are two reasons why a many-chip SSD architecture suffers from utilization related problems and idleness as the number of chips increases.
The first reason is parallelism dependency exhibited by an I/O request access pattern. Unlike the main memory systems, SSD request sizes vary from a few bytes to KBs (or MBs), and their data offsets can vary significantly, which in turn introduces unbalanced chip utilization and low parallelism at a system level. 
For example, some requests do not span across all internal resources, and some other requests keep pending due to chip-level conflicts, which in turn influences the number of resources that can be allocated at a given time. Therefore, the degree of internal parallelism that can be enjoyed depends highly on incoming I/O access patterns, referred to as \emph{parallelism dependency} in this work.
Another reason behind the utilization and idleness problems of emerging many-chip SSDs is low flash-level \emph{transactional-locality}. The transactional-locality in this work corresponds to the ability of the references that form a flash transaction to exhibit high flash-level parallelism. Since multiple dies and planes are connected to shared voltage drivers and a single multiplexed interface, flash-level resources need to be managed carefully in order to exploit the maximum amount of parallelism. Specifically, highly parallel data accesses across internal flash resources can \emph{only} be achieved when incoming requests span all of them (spatial) and all types of request transactions can be identified within a very short time period like a few cycles (temporal). As a result, low flash-level transactional-locality may introduce poor data access concurrency and extra idleness.

\begin{figure}
\centering
\includegraphics[width=1\linewidth, bb= 0 0 179 147]{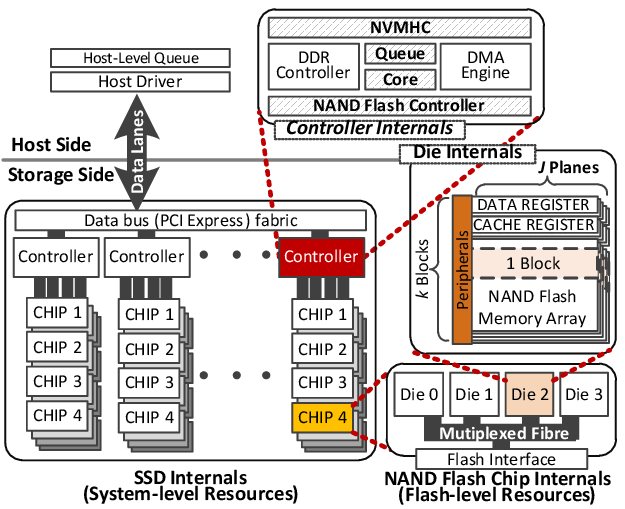}
\vspace{-10pt}
\caption{A many-chip SSD architecture. \label{fig:many-chip-arch}}
\vspace{-10pt}
\end{figure}

In this paper, we propose \emph{Sprinkler}, a novel device-level SSD controller, which targets maximizing resource utilization and achieving high performance without additional NAND flash chips. 
Specifically, Sprinkler relaxes the parallelism dependency by scheduling I/O requests based on \emph{internal resource layout} rather than the order imposed by the device-level queue (which is the current norm \cite{sched:ssdaware, pas, o3, pas}).
In addition, Sprinkler improves flash-level parallelism and reduces the number of transactions by \emph{over-committing} flash memory requests to specific internal resources. To the best of our knowledge, this is the first paper that suggests to exploit internal resource layout and over-commit flash memory requests in order to maximize resource utilization and parallelism, thereby improving many-chip SSD performance. Our main \textbf{contributions} can be summarized as follows:

\noindent \textbullet \emph{ Resource-driven I/O scheduling.} Unlike conventional SSD controllers which schedule flash memory requests based on the ``order of incoming I/O requests'', Sprinkler schedules them based on ``available physical flash resources'' in a fine-grain, out-of-order fashion. This method, called \emph{\textbf{\underline{R}}esource-driven \textbf{\underline{I}}/\textbf{\underline{O}} \textbf{\underline{S}}cheduling (RIOS)}, ``decouples'' parallelism dependency from the I/O request access patterns, timings and sizes, improving overall resource utilization by about 68.8\% under a wide variety of realistic workloads.  

\begin{figure*}
\centering
\includegraphics[width=\linewidth, bb=0 0 265 77]{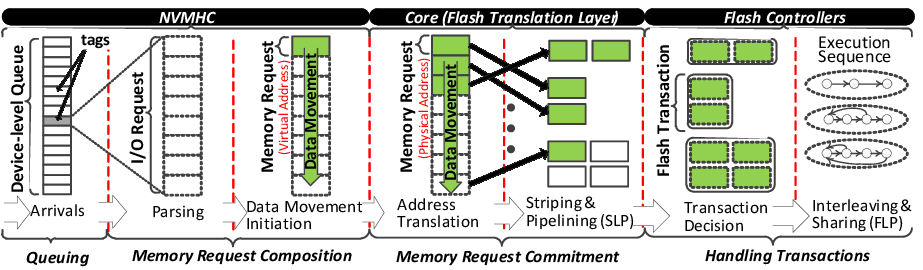}
\vspace{-5pt}
\captions{The operation of an I/O service routine in a many-chip SSD. In this routine, an I/O request is served by multiple phases and by different SSD internal resources in a pipelined fashion. \label{fig:io-process}}
\vspace{-10pt}
\end{figure*}

\noindent \textbullet \emph{ Flash memory request over-commitment.}  In order to increase flash-level transactional-locality, Sprinkler over-commits flash memory requests to devices. This \emph{\textbf{\underline{F}}lash-level parallelism \textbf{\underline{A}}ware \textbf{\underline{R}}equest \textbf{\underline{O}}ver-commitment (FARO)} maximizes the opportunities for building a flash transaction at runtime, parallelizing multiple memory requests with respect to the constraints imposed by the underlying flash technology. FARO provides, on average, 80.2\% more flash-level parallelism and reduces approximately 50\% of flash transactions.     

\noindent \textbullet \emph{ Reducing idleness in many-chip SSDs.} 
We identify two different types of idleness in an SSD: \emph{inter-chip idleness} and \emph{intra-chip idleness}. Sprinkler reduces inter-chip and intra-chip idleness by 46.1\% and 23.5\%, respectively. As compared to a conventional virtual address based scheduler \cite{sched:ssdaware, pas} and even a state-of-the-art physical address based scheduler \cite{o3, paq}, Sprinkler (RIOS and FARO together) provides at least 1.8 times better system throughput, and 78\% lower I/O latency.

\section{Many-Chip Solid State Disks}
Modern SSDs employ hundreds to thousands of flash chips by interconnecting multiple I/O channels and control units in an attempt to fill the performance gap between high speed interfaces (16GB/sec) and flash medium (40$\sim$400MB/sec). 
Figure \ref{fig:many-chip-arch} pictorially illustrates a recent many-chip SSD architecture by Mavell \cite{spec:mavell}. 
In this architecture, I/O services are dealt with a multi-phase approach carried out by different internal resources to take advantage of the overlap between I/O and computation, and exploit different levels of parallelism. 


\subsection{System-Level Technologies}
\noindent \textbf{Resources.} 
As shown in Figure \ref{fig:many-chip-arch}, there exist four main shared resources:

\noindent 1) \emph{Non-Volatile Memory Host Controller (NVMHC)} is a control logic, responsible for communication between the host and SSD internal components. Since NVMHC is the only component that is aware of the host interface protocols related to queueing, handshaking and data packet handling, device-level I/O schedulers are usually implemented in NVMHC.
 
\noindent 2) \emph{Core} is a microprocessor, which is dedicated to translate virtual addresses, compatible with host file systems block addresses, to flash memory physical addresses. To manage this address translation, an embedded software module, called \emph{Flash Translation Layer} (FTL), is implemented in the core. 

\noindent 3) \emph{Flash Controller} builds a \emph{flash transaction} from multiple memory requests, and executes it. 

\noindent 4) \emph{Channel} is a data path between the flash controller and the flash medium. Multiple flash chips are connected to a channel like an array and share the same path for data transfer between the SSD internal buffer and the flash chip internal registers.

\noindent \textbf{I/O Service Routine.}
As shown in Figure \ref{fig:io-process}, at the beginning of the process of servicing an I/O, NVMHC enqueues the host request information, called \emph{tags}, in its own device-level queue and schedules tags. Once a host \emph{I/O request} is chosen to be serviced, NVMHC parses the associated tag based on the underlying protocol (e.g., NVM Express \cite{spec:nvme}). NVMHC then builds a \emph{memory request} whose data size is the same as the atomic flash I/O unit size, and initiates the corresponding data movement between the host and the SSD. This activity is referred to as \emph{memory request composition}. Since, unlike the main memory systems, the length of an I/O request can vary significantly, ranging from several bytes to an MB, an I/O request is typically split into \emph{multiple} memory requests. 

In the next step, NVMHC sends these \emph{memory requests} to the core processor in the SSD for further I/O processing. This \emph{memory request commitment} needs to be performed in a timely fashion so that the multiple phases related to queuing, memory request composition and commitment can be pipelined.   
FTL translates virtual addresses of the memory requests coming from NVMHC into physical addresses, and scatters them over available flash controllers.
Finally, flash controllers build a \emph{flash transaction} by \emph{coalescing} multiple memory requests. The amount of parallelism exhibited by a flash transaction varies depending on the transactional-locality as well as the limitations of the underlying flash chip technology, as explained below.

\noindent \textbf{System-Level Parallelism (SLP).}
While serving an I/O request, many associated memory requests can be scattered across multiple internal resources. First, after the physical addresses have been determined by FTL, the corresponding accesses can be parallelized over multiple flash controllers and channels, and this process is called \emph{channel stripping}.  Further, each flash controller can pipeline a series of I/O commands, control commands and data movements, associated with the transaction across multiple flash chips within a channel, and this process is referred to as \emph{channel pipelining}. 
Even though channel stripping and channel pipelining aim to improve internal parallelism, the amount of SLP brought by them depends highly on the incoming I/O access pattern. As shown in Figure \ref{fig:small-read-utilidle}, poor chip utilization, caused mainly by parallelism dependency, is one of main reasons for performance stagnation on emerging many-chip SSDs.

\subsection{Flash-Level Technologies}
\noindent \textbf{Resources.}
State-of-the-art SSDs also employ the following flash-level technologies: 

\noindent 1) \emph{ Flash Chip} consists of multiple dies and planes, and exposes them through a small number of I/O (e.g., 8 \textasciitilde 16) pins and a CE (chip enable) pin to the system-level resources. This flash interface reduces the I/O connection complexity and communication noise, but it also introduces a set of command sequences and data movements for handing flash transactions. 

\noindent 2) \emph{ Die} is a memory island, connected to a single multiplexed bus fibre through the flash interface and the CE. Note that the memory cells in different dies can operate independently.  

\noindent 3) \emph{ Plane} is the memory array in a die, sharing the wordline and voltage drivers for accessing specific flash memory cells.

\noindent \textbf{Flash-Level Parallelism (FLP).} 
Even though multiple dies are squeezed into a single flash interface, a set of bus activities for a transaction (e.g, flash command, data movement, control signals) can be interlaced, and the multiple dies can independently work without any circuit-level modification. Consequently, multiple memory requests can be interleaved across dies via \emph{die interleaving}, which in turn improves chip throughput and response time for a transaction $n$ times, where $n$ is the number of flash dies. \emph{Plane sharing} activates multiple planes in a die through the shared wordline access, thereby improving throughput by $m$ times, $m$ being the number of planes. Lastly, die interleaving and plane sharing can be combined, which can improve transaction performance by approximately $n*m$ times. 

\noindent \textbf{Flash Transaction and Parallelism Dependency.}
Unlike DRAM, NAND flashes have an wide spectrum of operation sets, commands and execution sequences -- most flash memory vendors at least offer ten flash operations, each of which typically has a different execution sequence. In this context, a flash transaction is a series activities that the flash controller has to manage in executing a flash operation. It is composed of a set of commands (i.e., flash, control, delimiter commands), data movements (i.e., contents, addresses, status information). In addition, during the execution stage, all memory requests in the transaction are required to follow an appropriate timing sequence that flash makers define, and each flash transaction has its own timing sequence. Therefore, as shown in the handling transactions part of Figure \ref{fig:io-process}, the type of a transaction should be decided within a short period of time before entering the execution sequence. Due to this, the parallelism of each transaction potentially has a dependency on I/O access pattern, length and arrival timing of incoming memory requests, which makes it difficult to achieve high levels of FLP. We refer to this as ``parallelism dependency'' and demonstrate specific examples exhibiting low FLP below.  

\noindent \textbf{Challenge.} 
First, in cases where multiple memory requests arrive at a flash controller in a time interval which is longer than the transaction type decision time of the controller, they will be served by separate transactions even though they could have been serviced, from a flash-level perspective, as a single transaction. This poor flash-level temporal transactional-locality can potentially contribute to low FLP. On the other hand, since only one flash transaction can occupy the shared interface, bus and flash medium at a time, once the transaction type is determined and the corresponding memory requests are initiated, other memory requests heading to the same chip should be \emph{stalled} until the shared flash resources are free. Lastly, to take advantage of plane sharing, addresses of the memory requests in a transaction should indicate the \emph{same} page and die offset in the flash chip, but \emph{different} block addresses (or plane addresses). As a result, low flash-level spatial transactional-locality can also  introduce low FLP and high intra-chip idleness. 



\section{I/O Scheduling in Modern Controllers}

The state-of-the-art I/O scheduling schemes in NVMHCs can be broadly classified as \emph{virtually address schedulers (VAS)} \cite{sched:ssdaware, park, pas} and \emph{physically address schedulers (PAS)} \cite{o3, paq}, in terms of the address type of memory requests that they operate on. In this section, we briefly explain these two schedulers and their drawbacks. 

\begin{figure}
\centering
\def\subfigcapskip{0pt}
\subfloat[System-level view   ]{\label{fig:vas-patch}\rotatebox{0}{\includegraphics[width=1\linewidth]{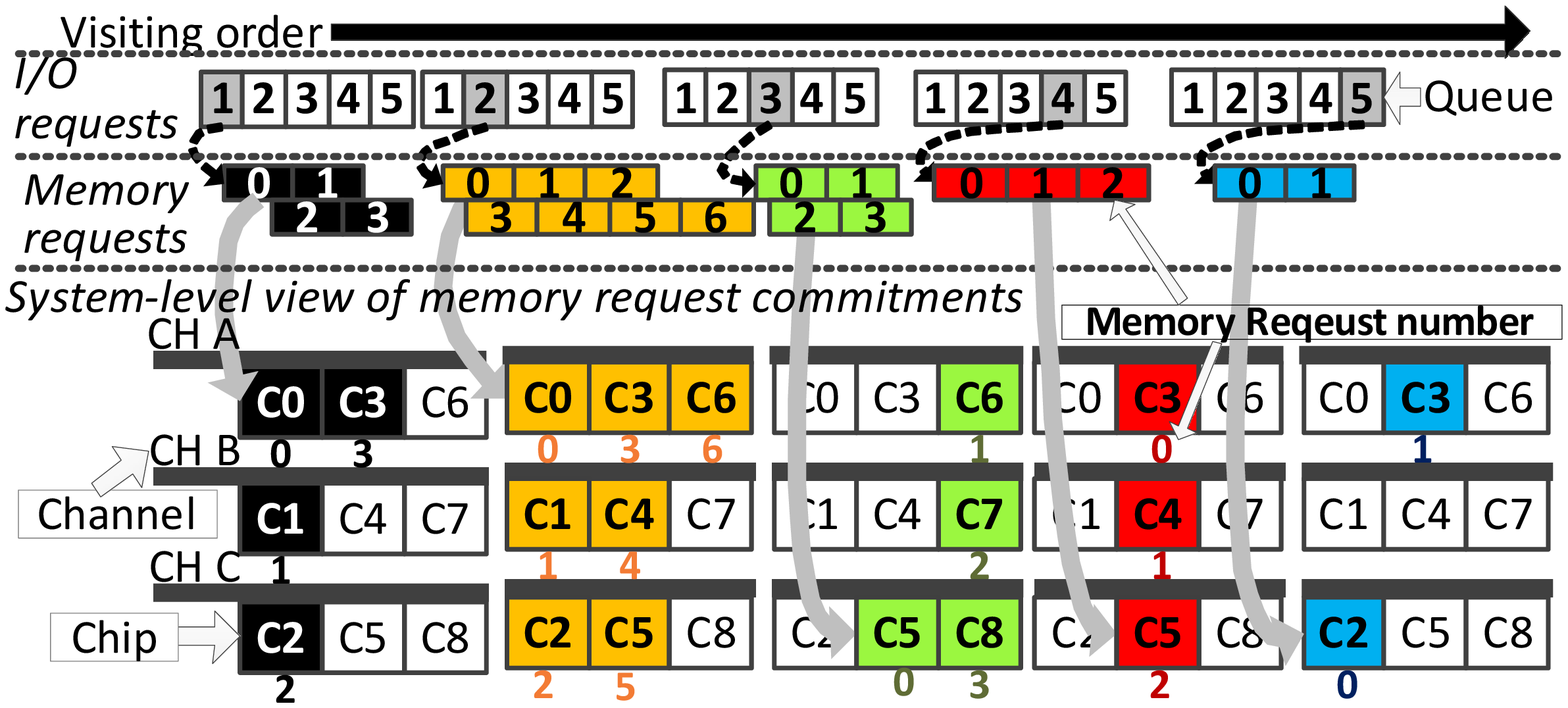}}}
\vspace{-10pt}

\subfloat[Flash-level view (VAS service time for chip 4 (C3))
]{\label{fig:vas-timing}\rotatebox{0}{\includegraphics[width=1\linewidth]{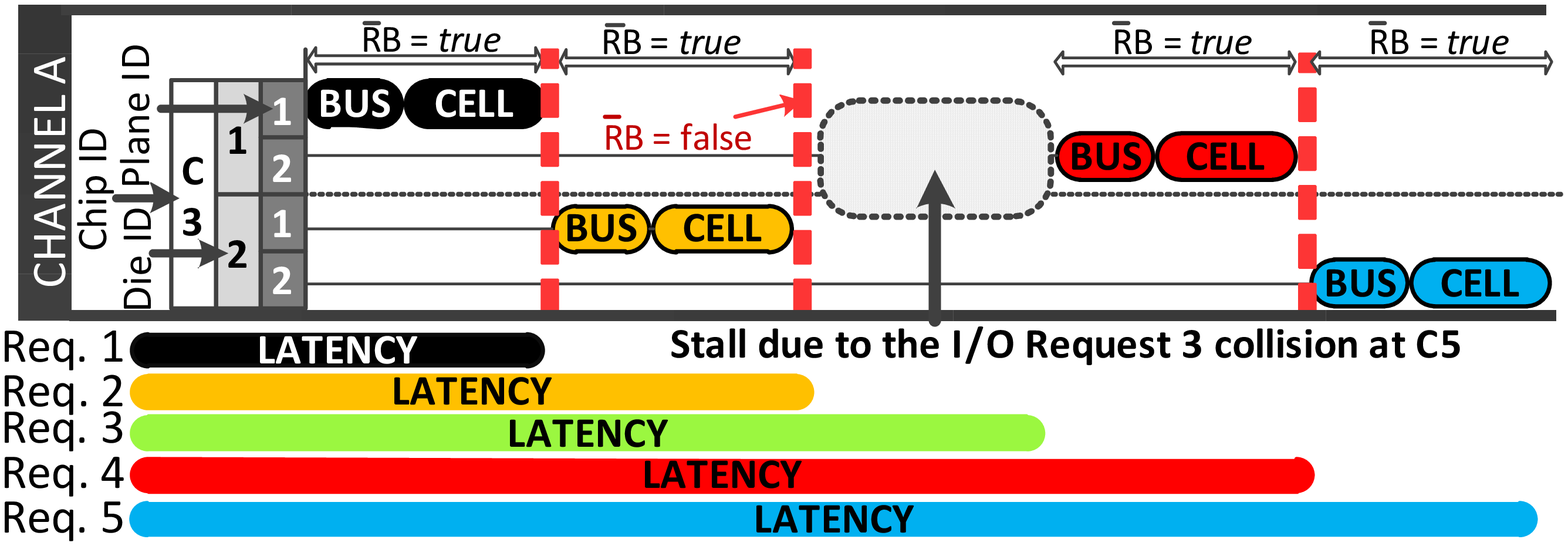}}}
\vspace{-5pt}
\captions{Operation of virtual address scheduler (VAS). VAS exhibits low resource utilization and long (I/O request) pending times at the flash level.}
\label{fig:vas}
\vspace{-10pt}
\end{figure}

\begin{figure}
\centering
\def\subfigcapskip{0pt}
\subfloat[System-level view  ]{\label{fig:pas-patch}\rotatebox{0}{\includegraphics[width=\linewidth]{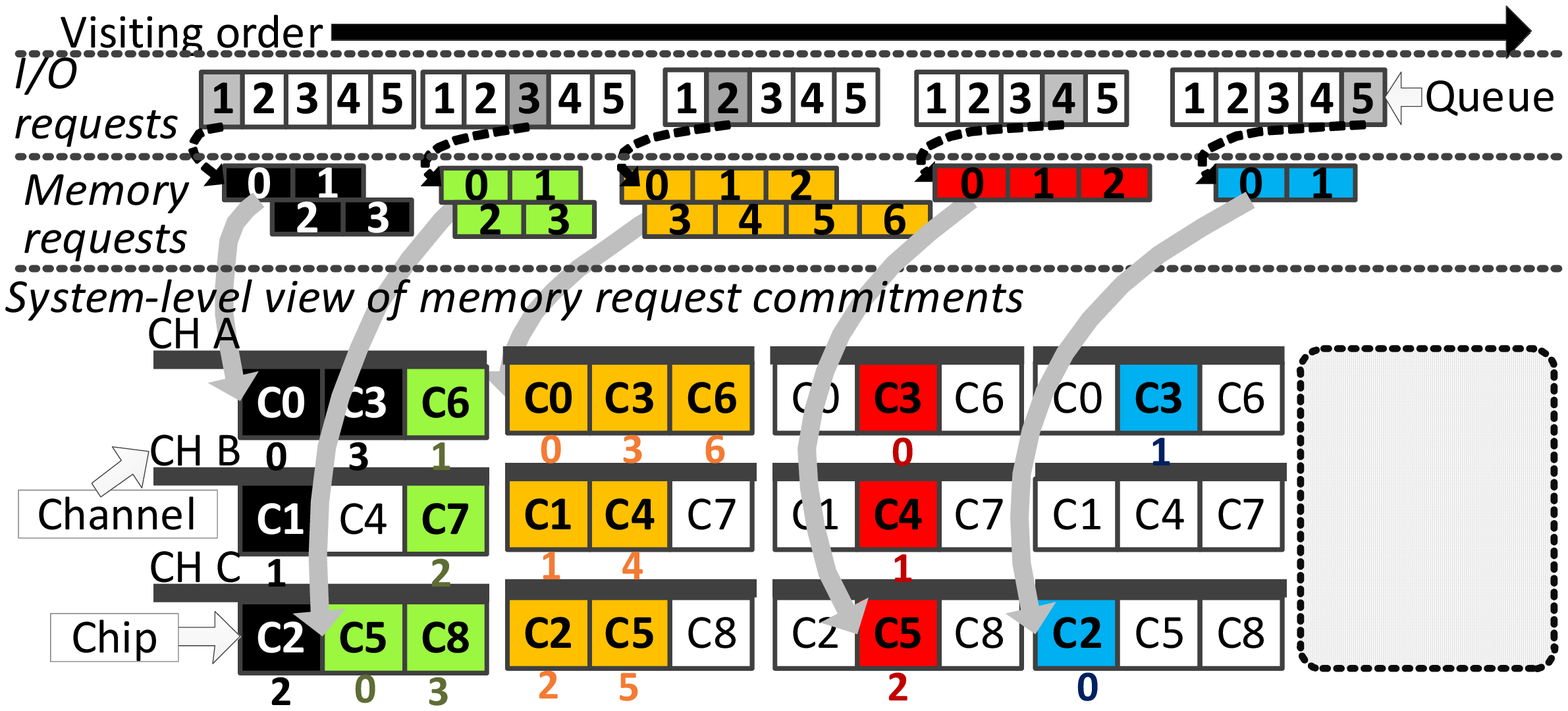}}}
\vspace{-10pt}

\subfloat[Flash-level view (PAS service time for memory requests)
]{\label{fig:pas-timing}\rotatebox{0}{\includegraphics[width=\linewidth]{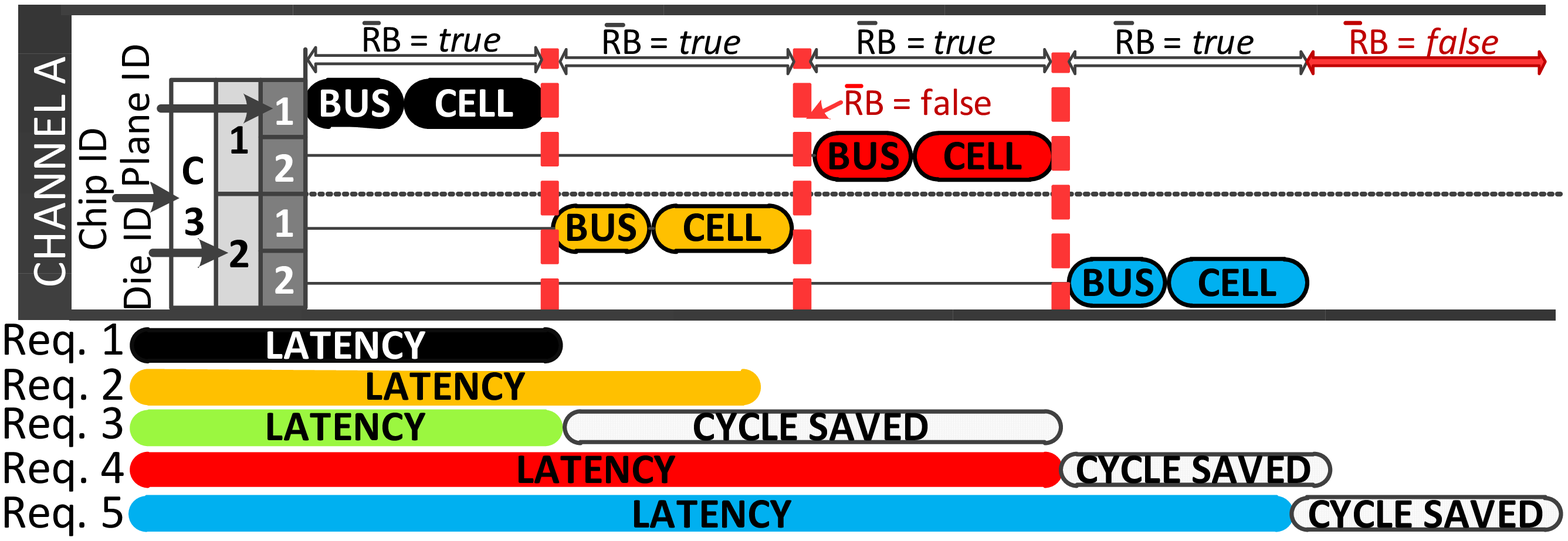}}}
\vspace{-5pt}
\captions{Operation of physical address scheduler (PAS). PAS exhibits better resource utilization and higher FLP than VAS, but it still suffers from parallelism dependency and low flash-level transactional-locality.}
\label{fig:pas}
\vspace{-10pt}
\end{figure}

\noindent \textbf{Virtually Address Scheduler (VAS).}
This type of scheduler decides the order of I/O requests in the device-level queue, but builds and commits memory requests relaying only on the virtual addresses of the I/O requests, provided by the underlying FTL. Consequently, VAS can suffer from collisions among I/O requests, which leads to low levels of SLP and FLP.   
Figure \ref{fig:vas-patch} illustrates this problem\footnotes{For all the system-level diagrams (Figures \ref{fig:vas-patch}, \ref{fig:pas-patch}, and \ref{fig:spk-patch}), we show snapshots with respect to the device-level queue (e.g., native command queue), memory request commitments, and the corresponding physical layout by following the controller visit order (from left to right). For each snapshot, an uncolored box on each channel indicates idle chips, and memory request numbers on the resource layout refer to the memory request commitment order.}. In this example, there exist five I/O requests arriving back-to-back and containing a total of nineteen individual memory requests. Initially, VAS composes four memory requests belonging to I/O (request) \#1 and strips/pipelines across four different chips (C1$\sim$C4). However, to commit I/O \#2 next, VAS has to wait for the completion of the previously-committed request, which makes five chips (C4$\sim$C8) idle. The reason behind this inter-chip idleness is the request collisions between I/O \#1 and I/O \#2 in three different chips (C0 \textasciitilde C2), and VAS schedules the I/Os with no idea about the underlying physical addresses.
Similarly, I/Os \#3, \#4, and \#5 would be stalled in the queue due to the request collisions with the previously-committed requests, which makes twenty chips idle. 

%

Figure \ref{fig:vas-timing} plots a microscopic view of C3 in this example\footnotes{For all flash-level view diagrams (Figure \ref{fig:vas-timing}, \ref{fig:pas-timing}, and \ref{fig:spk-timing}), we show bus and cell timing diagram associated with each system-level view's snapshot (from left to right). To make better comparisons, we also illustrate the corresponding I/O request level latency below them. Note that all the schedulers we discuss can only submit a flash transaction during  the \={R}B=$false$ periods and have the same type of out-of-order executable device level queue (NCQ).}. 
Since VAS has no other request commitments at the beginning of the I/O process, the first transaction is built by only considering I/O \#1. When C3 in serving the flash transaction, it makes read/busy signal ($\bar{R}B$) true, which means that the chip is not available to serve anything else. 
Even though VAS is ready to commit the memory request associated with I/O \#2 as the next step, it has to wait until $\bar{R}B$ becomes false.
Similarly, since VAS has no knowledge about the underlying physical layout, it further commits three memory requests heading to C3 (associated with 3 different I/Os: \#2, \#4 and \#5) in tandem without any transactional-locality consideration. Consequently, they are built as four different ``flash transactions'' at the chip-level including one stalled time frame introduced by I/O  \#3.
One can see from this example that \emph{we have a large scope for improving utilization by reducing the number of idle chips}, if VAS could reorder the I/O requests by being aware of ``physical addresses''.

\noindent \textbf{Physically Address Scheduler (PAS).}
PAS schedules the I/O requests by being aware of the physical addresses exposed by a hardware-assisted preprocessor \cite{o3} or a software-based address translation unit \cite{paq}. Thanks to this physical address space exposure, PAS can reorder I/O requests in an attempt to address the request collision problem and execute flash transactions in an out-of-order fashion. 
Specifically, from a system-level viewpoint, PAS can serve multiple I/O requests by grouping them without a major memory request collision, which in turn reduces the number of idle chips and improves SLP. 
Figure \ref{fig:pas-patch} plots this scheduling scenario. 
PAS swaps I/O \#3 with I/O \#2 since the latter can be simultaneously executed with I/O \#1 without any shared resource conflict. 
Due to the high degree of SLP at the beginning of the commitment process, the stalled time frame shown in Figure \ref{fig:vas-timing} can be eliminated, and I/Os \#3, \#4 and \#5 save multiple execution cycles, which impacts both system latency and throughput. Since this I/O reordering scheme based on physical addresses can partially relax parallelism dependency, there are only fifteen idle chips in order to complete all the I/O requests in the queue, which corresponds to a 40\% reduction in inter-chip idleness compared to VAS (Figure \ref{fig:vas-patch}).

However, PAS still has two downsides. First, it composes memory requests and commits them based on ``I/O request arrival order'', and these arrival patterns can vary (in terms of length and data offset of I/O requests). Second, PAS cannot capitalize on flash-level spatial transactional-locality even if one has lots of enqueued I/O requests, because \emph{it does not take into account the   physical layout} of underlying resources, chips, and microarchitecture configurations. In addition, memory request commitments performed by PAS, clueless about flash-level temporal transactional-locality, can introduce poor FLP and long intra-chip idleness. 
Figure \ref{fig:pas-timing} plots this problem for C3 from a flash level perspective. PAS serves multiple I/O requests in parallel, and shortens their latencies as shown in the bottom of the figure. However, I/O \#4 and I/O \#5 still need to be stalled due to the resource collision imposed by the host-level request information and boundary limit, and therefore, each memory request is assigned to a different transaction, which in turn introduces long latencies for I/O \#4 and I/O \#5. This example provides two insights. First, \emph{the total number of chips is relatively fewer than the total number of memory requests} coming from different I/O requests, which means that many flash chips can be activated at any given time if parallelism dependency could be relaxed.
Second, \emph{there exist (at any given period of time) many requests heading to the same chip but to different internal resources}, which implies that multiple memory requests can be built into an FLP transaction if we could change their commitment order.

\begin{figure}
\centering
\includegraphics[scale=0.25]{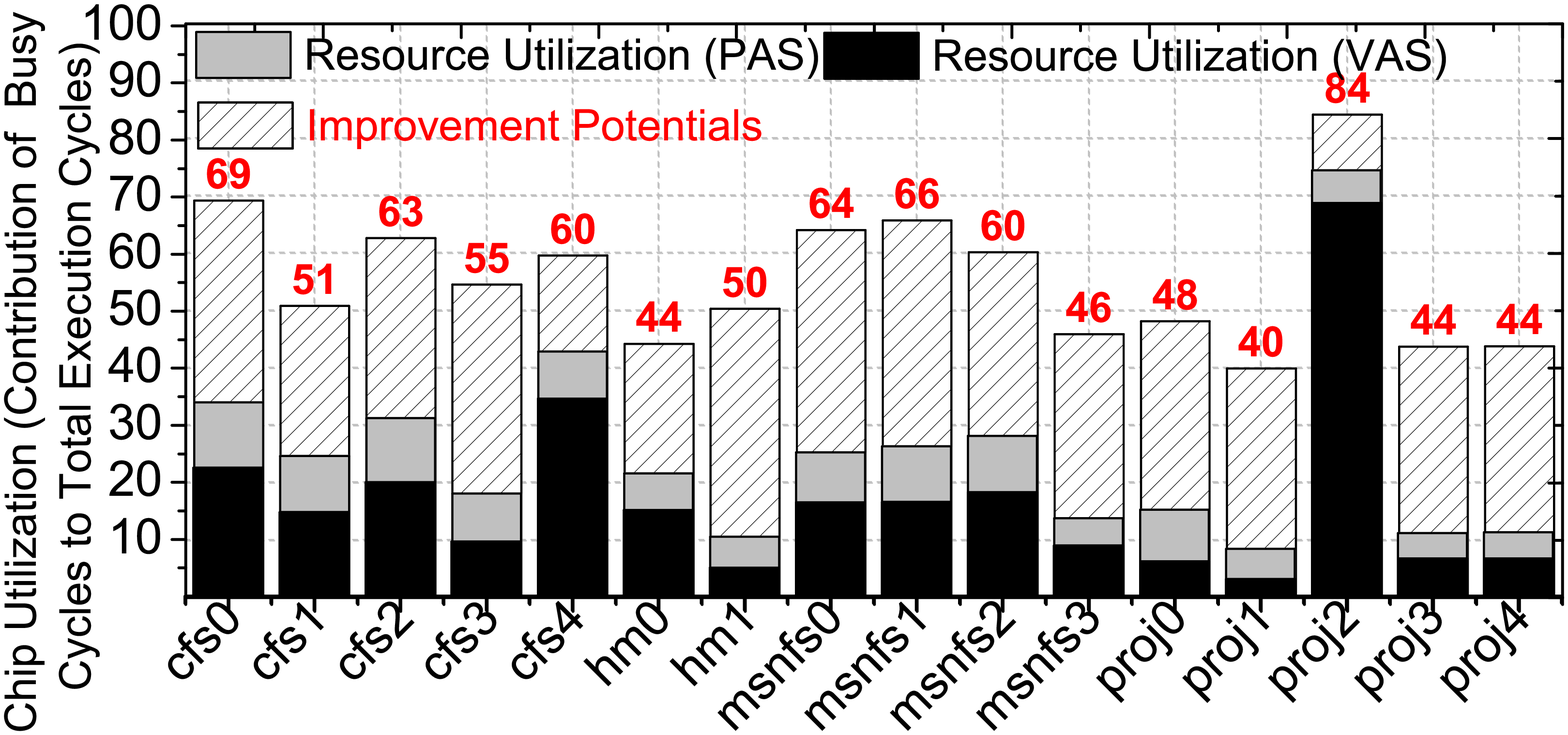}
\vspace{-10pt}
\captions{Resource utilization and improvement potential under various workloads. Relaxing parallelism dependency and achieving high transactional-locality improve resource utilization by 3x and 2x compared to VAS and PAS, respectively. 
\vspace{-10pt}
\label{fig:imp-poten}}
\end{figure}

\noindent \textbf{Improvement Potential.}
Strong parallelism dependency and low flash-level spatial/temporal transactional-locality introduce poor internal resource utilization that prevents a many-chip SSD architecture from realizing its potential.
To quantify the potential gains when these two main challenges are removed, we simulated the same SSD platform used to  collect the data presented in Figure \ref{fig:many-chip-motive}, under sixteen publicly-available workloads \cite{exp:msr, exp:iotta}. 
Figure \ref{fig:imp-poten} plots the chip utilization exhibited by a state-of-the-art SSD controller under three different scenarios: 1) A typical scenario where these two challenges are present, indicated by the black portion in each bar, 2) An improved scenario where resource conflicts are addressed; captured by the gray portion in each bar, and  3) A scenario where the parallelism dependency is fully relaxed and high transactional-locality is guaranteed, which is indicated by the shaded portion in each bar. 
It can be observed from this plot that, internal resources are badly underutilized under the first (typical) scenario. Specifically, we observe an average chip utilization of 17\% for the typical case scenario (VAS) and 24\% for the improved scenario (PAS).  
However, in cases where the two challenges mentioned above are eliminated, all observed resource utilizations are over 40\%, irrespective of the workload access pattern. 
Specifically, relaxing parallelism dependency and achieving high transactional-locality improve resource utilization by 3x and 2x, respectively, compared to the typical scenario and improved scenario, and the chip utilization reaches in this case 55\%, on average. Overall, \emph{these results clearly underline the importance of addressing parallelism dependency and transactional-locality} in SSDs. Next, we present our proposed scheduling strategy that addresses these challenges.

\section{Sprinkler}
To maximize resource utilization and reduce idleness, we propose Sprinkler, which is a novel scheduling strategy composed of two components: 1) \textbf{RIOS} and 2) \textbf{FARO}.
Unlike VAS and PAS, which consider the I/O request order in the device-level queue and build memory requests based on host level information, RIOS schedules and builds memory requests based on the ``internal resource layout'' to fully \emph{relax} the parallelism dependency. The relaxed parallelism dependency through RIOS leads to the activation of as many system-level and flash-level resources as possible with minimal impact of the length  and data offset of the incoming I/O requests. 
Further, FARO over-commits flash memory requests with the goal of achieving high  temporal and spatial transactional-locality, which can in turn significantly improve FLP and reduce intra-chip idleness. Specifically, FARO supplies many memory requests to the underlying flash controllers, which can increase the opportunities for building a high-FLP transaction at a chip level. In addition, the over-committed memory requests by FARO satisfy appropriate timing sequences, bringing higher temporal transactional-locality as well (more on this later). Due to this improved transactional-locality, under  FARO, all memory requests targeting toward the same chip but different die and plane can be incarnated as a ``single'' flash transaction.  

%

\subsection{Resource-Driven I/O Scheduling}
\label{sec:rios}

\begin{figure}
\centering
\includegraphics[width=1\linewidth]{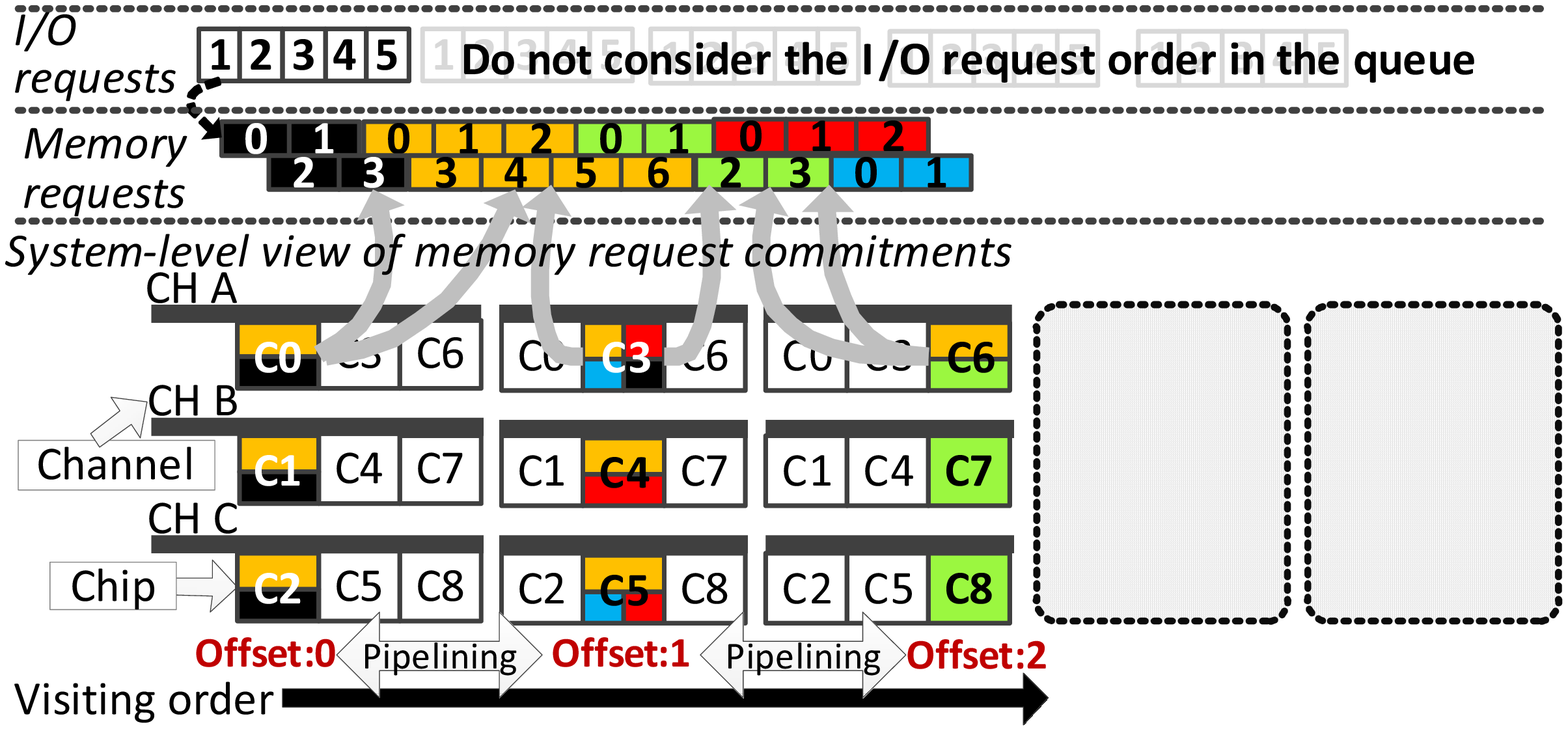}
\vspace{-5pt}
\captions{Operations of RIOS.  Sprinkler executes memory requests in a fine grain out-of-order fashion through its resource-driven I/O scheduling (RIOS) strategy. In this example, Sprinkler can eliminate one and two more snapshots (indicated by dashed lines) on the visit order time-line compared to Figures \ref{fig:vas-patch} and \ref{fig:pas-patch}. \label{fig:spk-patch}}
\vspace{-10pt}
\end{figure}

One of the insights behind Sprinkler is that the stalled memory requests in the queue can be immediately served, if the scheduler could compose the requests beyond the boundary of host-level I/O requests and commit them regardless of the order of the I/O requests. 
Motivated by this, Sprinkler does \emph{not} consider the I/O request order in the queue and does \emph{not} build flash transaction based on incoming host information. 
Instead, it schedules memory requests based on the physical resource information, composed by the following three processes. (i) Securing tags without actual data movement until there is no more room to enqueue or no further back-to-back I/Os from a host. 
In this step, the scheduler identifies the resources targeted by I/Os and logically categorizes the requests per physical chip without any memory request composition. (ii) Initiating data movement, composing memory requests and committing them per flash chip \emph{not} per I/O, by traversing all flash chips in the SSD. During this process, the necessary data movements between the host and the SSD are performed in an out-of-order fashion. Further, the computation in step $i$ can be overlapped with the data movement in step $ii$. (iii) Keep continuing with step $ii$ until the queue secures an available room. 
We refer to this scheduling strategy as resource-driven I/O scheduling (RIOS), which can be viewed as a type of \emph{fine-grain out-of-order execution strategy}. This parallelism dependency relaxation allows RIOS maximize the number of active flash chips at any given time, irrespective of the I/O access pattern observed. 

One potential problem with RIOS is that, if it commits memory requests by visiting flash chips in an arbitrary fashion, it can introduce undesirable ``system level'' resource contention. For instance, if RIOS visits each flash chip in a channel-first fashion (e.g., C0, C3, C6 in the example of Figure \ref{fig:pas-patch}), the bus activities of each memory request such as flash commands, control commands and data movements, require channel bus arbitration, which leads to I/O serialization to some extent. To avoid this, RIOS visits the flash chips that have the same offset in each channel, across different channels. It then increases the chip offset, and continues to visit the corresponding flash chips for memory request composition and commitment until all flash chips in the SSD are visited. This traversal order allows RIOS scatter memory requests across the system, taking advantage of channel stripping and channel pipelining so that their computation activities can be overlapped with I/O, and their memory requests can be fully parallelized without any major resource conflicts.

Figure \ref{fig:spk-patch} illustrates how differently RIOS schedules the five I/Os shown, compared to VAS (Figure \ref{fig:vas-patch}) and PAS (Figure \ref{fig:pas-patch}). 
From the beginning of the I/O scheduling, RIOS composes six memory requests associated with I/O \#1 and I/O \#2, and commits them to C0, C1, and C2, whose chip offset is zero. While data movements corresponding to I/O \#1 and I/O \#2 are being performed, RIOS in parallel composes eight memory requests and commits them to C3, C4 and C5 by increasing the chip offset. Lastly, RIOS schedules four memory requests related to I/O \#4 and I/O \#5 to the remaining chips, namely, C6, C7 and C8. 
Since all these individual steps can be pipelined (they do not have the same chip offsets), RIOS significantly reduces inter-chip idleness by relaxing the parallelism dependency and presents opportunities for building transactions with high degree of FLP.

\begin{figure}
\centering
\includegraphics[width=1\linewidth]{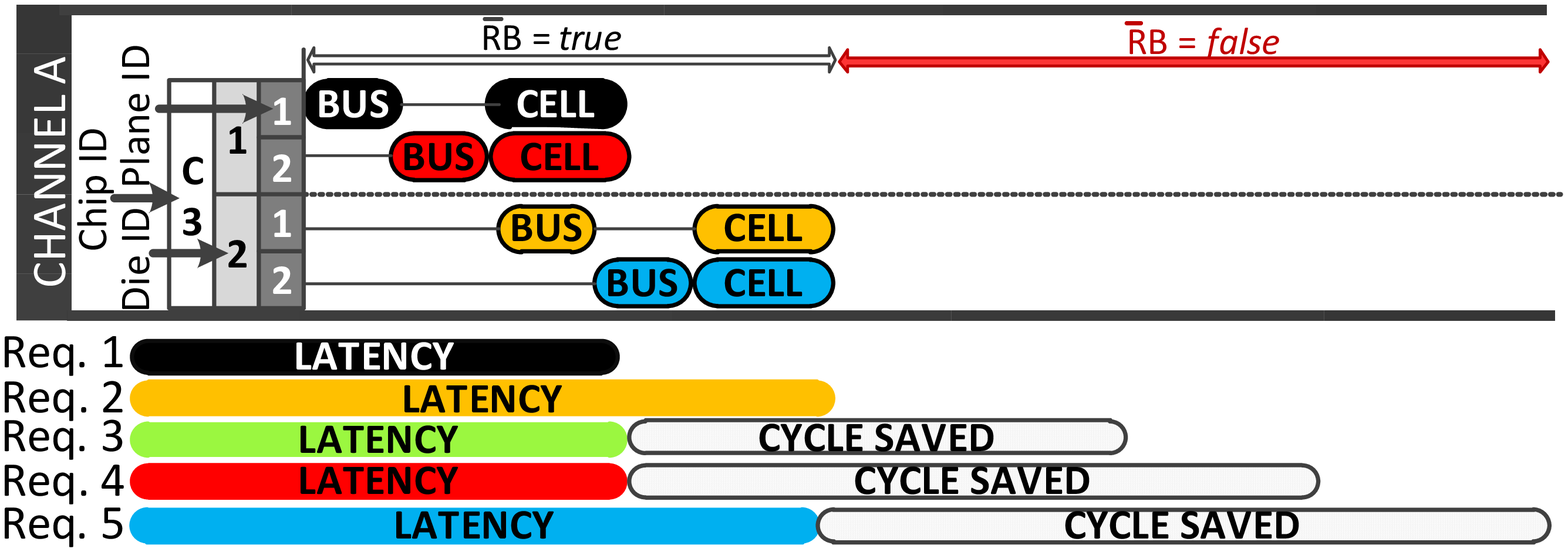}
\vspace{-5pt}
\captions{FARO service timing diagram. Four memory requests can be served as a die interleaving with multiplane transaction \cite{arch:dtssd, pas}, which only takes four bus activities and a cell activity from a system level viewpoint. Note that this flash transaction composition is not a part of our scheduling method, which means that VAS and PAS could also achieve this transaction if they could address parallelism dependency and appropriately schedule them by being aware of the underlying physical resource layout.}   
\label{fig:spk-timing}
\vspace{-10pt}
\end{figure}

\subsection{FLP-Aware Memory Request Overcommitment}
\label{sec:faro}

The scheduling strategy adopted by NVMHC can assist flash controllers to build high-FLP transactions by being aware of the underlying flash microarchitecture characteristics. In order to improve the opportunities for building high-FLP transactions at runtime, Sprinkler also \emph{over-commits} memory requests for each chip; this is referred to as FLP-aware memory request overcommitment (FARO). 
The idea behind FARO is that, if the scheduler is able to early-commit multiple memory requests targeting different flash internal resources in a chip, it can have more flexibility in building a flash transaction with high parallelism at the very beginning of the flash command handling. 
Motivated by this, our proposed FARO brings as many requests as possible to flash controllers as early as possible, allowing them to coalesce multiple memory requests into a single flash transaction with better flash-level spatial and temporal transactional-locality. 
Figure \ref{fig:spk-timing} illustrates how much the over-committed memory requests (by FARO) can shorten the latencies of different I/O requests, compared to VAS (Figure \ref{fig:vas-timing}) and PAS (Figure \ref{fig:pas-timing}). 
With FARO, four over-committed memory requests associated with four different I/Os (\#1, \#2, \#4 and \#5) present high transactional-locality at the time of building a flash transaction, and therefore, they can be built (using die interleaving and plane sharing FLP) as a single transaction, which represents the highest FLP as far as  C3 is concerned. Consequently, for the four memory requests shown, system-level resources experience only four bus activities and the single flash memory cell activity. In this example, FARO saves more than half of the I/O execution latencies for I/O requests \#3, \#4, and \#5 in the queue. The number of cycles saved by FARO is shown in the bottom of the figure.

\begin{figure}
\centering
\includegraphics[width=1\linewidth, bb=0 0 188 74]{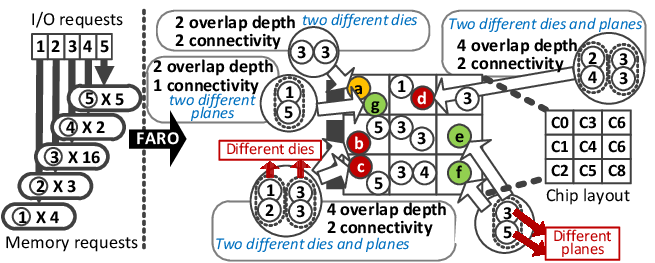}
\vspace{-5pt}
\captions{FLP-Aware Request Overcommitment (FARO). FARO increases transactional-locality by controlling the overcommitment priority dynamically. Note that all the chips (on the chip layout) are involved in I/O executions, and many of them have transactions consist of multiple memory requests coming from different I/O requests across multiple queue entries.
\label{fig:faro}}
\vspace{-10pt}
\end{figure}

One potential concern with FARO is that it might increase, in certain cases, the flash-level resource contention if it over-commits the memory requests without any preference. To address this potential problem, our implementation of FARO considers \emph{overlap depth} and \emph{connectivity} among multiple memory requests in an attempt to control the overcommitment priority \emph{dynamically}. 
Overlap depth is the number of memory requests targeting different planes and dies in the same flash chip. 
Connectivity on the other hand is the maximum number of memory requests that belong to the same I/O request. 
While the overlap depth is a metric oriented towards improving FLP, the connectivity is a metric that targets improving I/O latency.   
Sprinkler gives the highest priority to the request with the highest overlap depth. In cases where there exist multiple requests having the same depth value, FARO over-commits the memory requests that have the highest connectivity among them. This dynamic priority control employed by FARO alleviates potential resource contention that could be caused by over-committed memory requests.
Figure \ref{fig:faro} explains how connectivity and overlap depth are used by FARO. In this example, the five I/Os have thirteen memory requests in total. Especially, the length (the number of memory requests) of I/O \#3 exceeds the total number of chips, which introduces a connectivity value of more than two in several chips (e.g., C1, C2, and C3). There exist twelve memory requests having an overlap depth of 4 targeting chips C1, C2, and C3. Since these memory requests head toward two different dies and planes, they can be coalesced as a single memory transaction (indicated by `b', `c' and `d' in the figure), and FARO commits them first, instead of the memory requests belonging to I/O \#5 or I/O \#1. 
In contrast, the memory requests heading to C0 can be captured by two transactions, `a' and `g'. 
Since `a' is composed of memory requests related to the same I/O \#3, FARO over-commits memory requests associated with `a', instead of requests \#1 and \#5 in an attempt to improve the latency of I/O \#3 as the second option. 

\subsection{Handling Live Data Migration}
One problem with the physical address schedulers implemented in NVMHC is the handling of live data migrations, which is the process of reading valid data pages, writing them into new locations, and updating the mapping information regarding physical addresses. It is critical to note that live data migrations can change physical addresses during an I/O service. Depending on the underlying flash firmware strategy, the reason why live data migration is invoked at runtime can be different, but usually the migration is performed because of, 1) garbage collection, 2) wear-leveling, or 3) bad block replacement, and the corresponding migration activities are similar in each case \cite{agcdgc, amp}. To address the migration problem, we introduce a \emph{readdressing callback strategy}, which is a conventional callback routine that updates the physical data layout information in the upper-level I/O scheduler. Since Sprinkler exploits the internal resource layout rather than the physical address of memory requests, our readdressing callback is invoked \emph{only if} the live data have been migrated between \emph{different flash internal resources}. 

\SetAlFnt{\scriptsize\sf}
\DecMargin{0.5em}
\begin{algorithm}
	\DontPrintSemicolon
	\SetAlgoVlined

	\tcc{Resource-Driven I/O Scheduling (RIOS)}
	\While{queue.$full$() != true || host.$next\_request$(\&tag) != null}
	{
		\tcc{queuing incoming I/O requests based on physical layout}
		tuple chip(chip\_idx, die\_idx, plane\_idx) := core.$preprocess$(tag)\;
		phy\_layout[chip\_idx].$insert$(chip, tag)\;
	}
	\tcc{try to enqueue tags as many as possible}
	\While{host.$next\_request$(\&tag) = null || queue.full() = true}
	{
		
		\For{i :=0 \emph{\KwTo}i/num\_channel}
		{
			\tcc{stripping}
			\For{j :=0 \emph{\KwTo}num\_channel}
			{
				\tcc{pipelining}
				\tcc{computation can be overlapped with flash I/O time}
				chip := phy\_layout.$get\_chip$(i*num\_channel + j)\;
				\tcc{FLP-Aware Request Over-commitment (FARO)}
				tag := $get\_highest\_overlap\_depth$(chip)\;
				\If{tag = null}
				{
					\tcc{there are tags which have same overlap depth}
					tag := $get\_tag\_considering\_connectivity$(chip)\;
				}
				mem\_vector := $build\_memory\_request$(tag)\;
				$commit\_mem\_requests$(mem\_vector)\;
			}
		}
	}

\setstretch{0.2} 
	\captions{ Sprinkle(host, queue, core) in NVMHC. 
	Note that necessary data movement initiations selectively occur, and composed memory requests are scheduled based on the physical layout information.}
	\label{alg:phy}
\vspace{-15pt} 
\end{algorithm}

\subsection{Implementation Details}
\noindent \textbf{Algorithm.} Algorithm \ref{alg:phy} describes how our scheduler sprinkles memory requests across multiple internal resources. Sprinkler first identifies the physical layout of the memory requests in terms of indexes of chip, die and plane. It keeps  queuing the incoming I/O requests and identifying their physical layouts until there is no parallel I/O request left or the queue has no more room. In the main scheduling part, Sprinkler selectively initiates data movements of memory requests targeting the currently-visited flash chip. In this step, Sprinkler visits flash chips following the traversal order explained in Section \ref{sec:rios}. It then checks which memory request has a high overlap depth and a high connectivity in order to identify the memory requests to  over-commit memory requests. During this scheduling step, Sprinkler keeps watching whether we have new I/O request arrivals and queue is full or not to maximize the fine grain out-of-order execution potential as well as to relax parallelism dependency.

\noindent \textbf{Complexity.} The memory space requirements to record the required physical layout information (resource identification) are negligible. Specifically, four bytes are sufficient to store each piece of information, including chip, die and plane indices, and we observed that a total of 256KB is sufficient to cover all the workloads tested in Section \ref{sec:eval}. In addition, the computation in each iteration can be overlapped with data movement between the host and the SSD. Note that \emph{any other scheduler implemented in NVMHC would have similar computation and space complexities to compose/commit requests}.

\noindent \textbf{The Order of Output Data.}
NVMHC maintains an eight byte memory request bitmap, which covers 128KB$\sim$1MB block size per queue entry. Each bit of this bitmap indicates an issued memory request. When the flash controller makes an upcall to inform a flash transaction completion, it clears bits corresponding to the memory requests associated with the transaction. The DMA engine brings back the data from the beginning of the I/O request offset to the host using multiple payloads in an in-order fashion. 
Note that \emph{this I/O completion process and bitmaps are required regardless of type of the scheduling strategy} implemented in NVMHC.

\noindent \textbf{Hazard Control.}
Sprinkler only schedules I/O requests using tags, which means that the actual data for writes sit on the host side buffer during the scheduling activity. Therefore, in the read-after-write and write-after-write cases, a host-side logical block adapter (or operating system) can simply return or overwrite the data using its own buffer. Because of this, in general, the standard storage-interface protocols do not specify any rules on data integrity and consistency management for request reordering. Nevertheless, we manage the write-after-read case by serving the read memory-requests first (\emph{only if} the target contents of reads are the same as the contents of writes in the plane-level) when FARO considers overlap-depth and connectivity, and serves I/Os without any reordering if there exists a force-unit-access command request, used by an OS to manage the integrity and consistency of the storage system.

\begin{table}
\centering
\vspace{-5pt}
\scriptsize
\begin{tabular}{|p{17pt}|l|l|l|l|l|l|p{20pt}|}
\hline
&\multicolumn{2}{p{40pt}|}{Total transfer size (MB)}& 	\multicolumn{2}{p{40pt}|}{Numbers of Instructions}& 	\multicolumn{2}{p{40pt}|}{Randomness (\%)}& Transac\newline -tional \\ 
\cline{1-7}
&Read& 	Write& 	Read& 	Write& 	Read& 	Write&locality \\ 
\hline
cfs0	&3607	&1692	&406	&135	&92.79	&86.59	& Low  \\
cfs1	&2955	&1773	&385	&130	&94.01	&86.12	&Medium   \\
cfs2	&2904	&1845	&384	&135	&94.28	&85.95	&Low   \\
cfs3	&3143	&1649	&387	&132	&93.97	&86.7	&High   \\
cfs4	&3600	&1660	&401	&132	&92.6	&86.59	&High   \\
hm0	&10445	&21471	&1417	&2575	&94.2	&92.84	&Medium \\ 
hm1	&8670	&567	&580	&28	&98.29	&98.59	&Medium   \\
msnfs0 	&1971	&30519	&41	&1467	&99.79	&87.23	&Low \\
msnfs1 	&17661	&17722	&121	&2100	&88.8	&66.71	&Low \\ 
msnfs2 	&92772	&24835	&9624	&3003	&98.13	&99.97	&High \\ 
msnfs3	&5	&2387	&1	&5	&22.52	&64.79	&High   \\
proj0	&9407	&151274	&527	&3697	&92.05	&79.31	&Medium  \\ 
proj1	&786810	&2496	&2496	&21142	&82.34	&96.88	&Medium   \\
proj2	&1065308	&176879	&25641	&3624	&78.74	&93.93	&Low \\
proj3	&19123	&2754	&2128	&116	&75.01	&88.37	&Medium   \\
proj4	&150604	&1058	&6369	&95	&84.39	&95.52	&Medium   \\

\hline
\end{tabular}
\vspace{-5pt}
\normalsize 
			\captions{
				Classifying our traces in terms of data transfer sizes, the number of I/O instructions, and randomness of the issued reads and writes. The last column gives transactional locality by statically analyzing the traces based on the chip layout we simulated in cases where there is no limit for the device-level queue size. 
			}
			\label{tab:workloads}
\vspace{-10pt}
\end{table}

\begin{figure*}[t]
\centering
\def\subfigcapskip{0pt}
\subfloat[I/O bandwidth  ]{\label{fig:band}\rotatebox{0}{\includegraphics[width=0.5\linewidth]{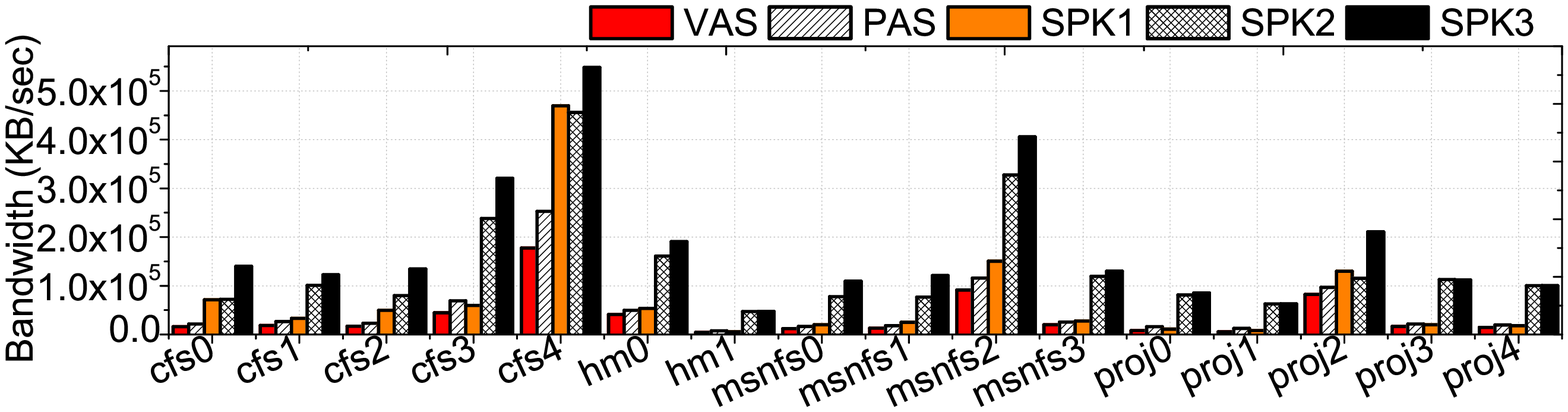}}}
\subfloat[IOPS(I/O operations per second)
]{\label{fig:iops}\rotatebox{0}{\includegraphics[width=0.5\linewidth]{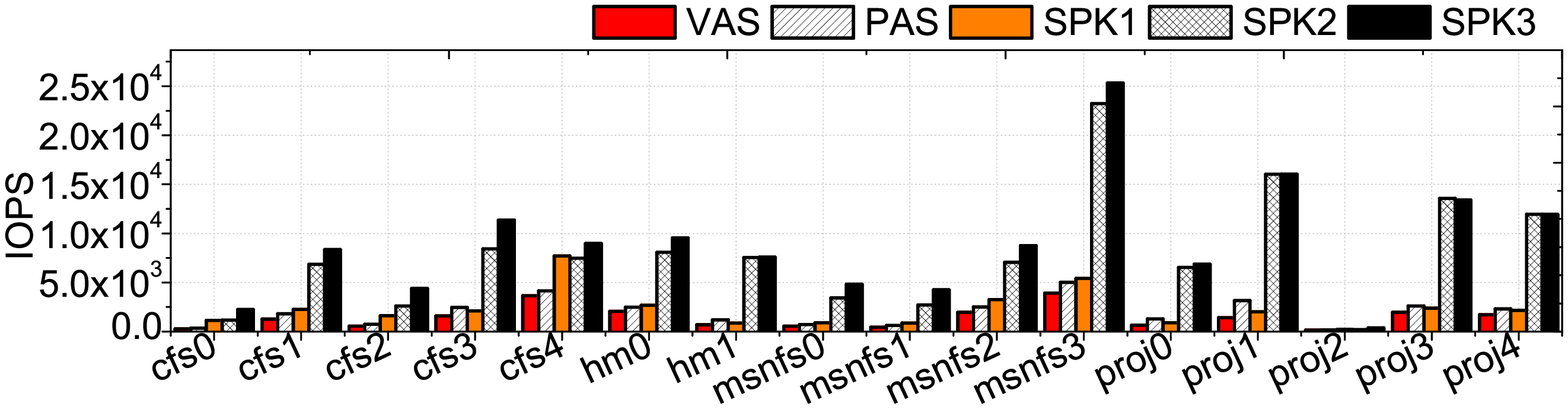}}}
\vspace{-5pt}

\subfloat[Average I/O latency  ]{\label{fig:avg-latency}\rotatebox{0}{\includegraphics[width=0.5\linewidth]{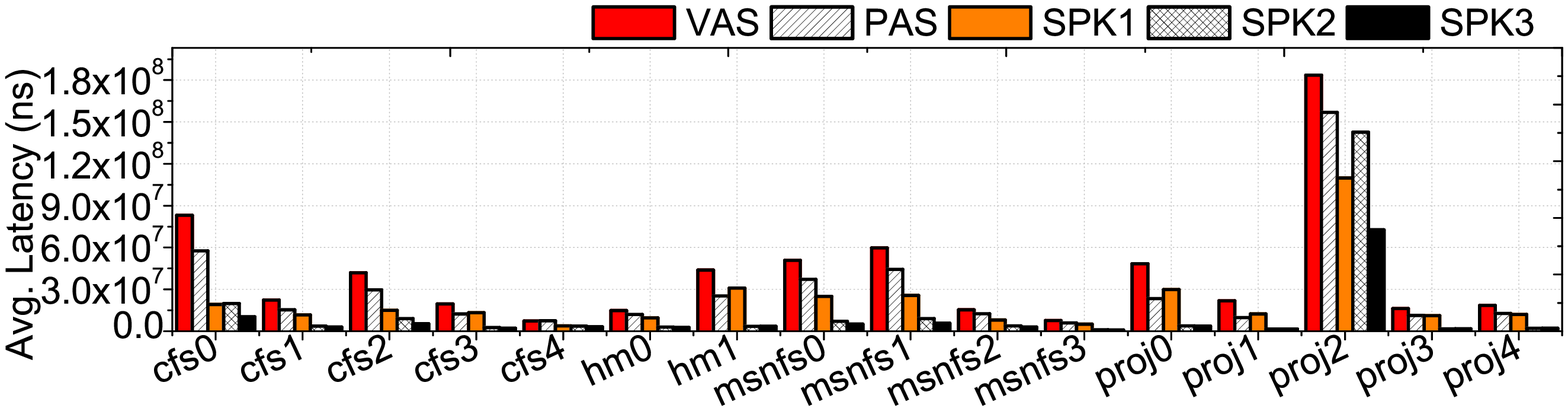}}}
\subfloat[Queue stall time.
]{\label{fig:pending}\rotatebox{0}{\includegraphics[width=0.5\linewidth]{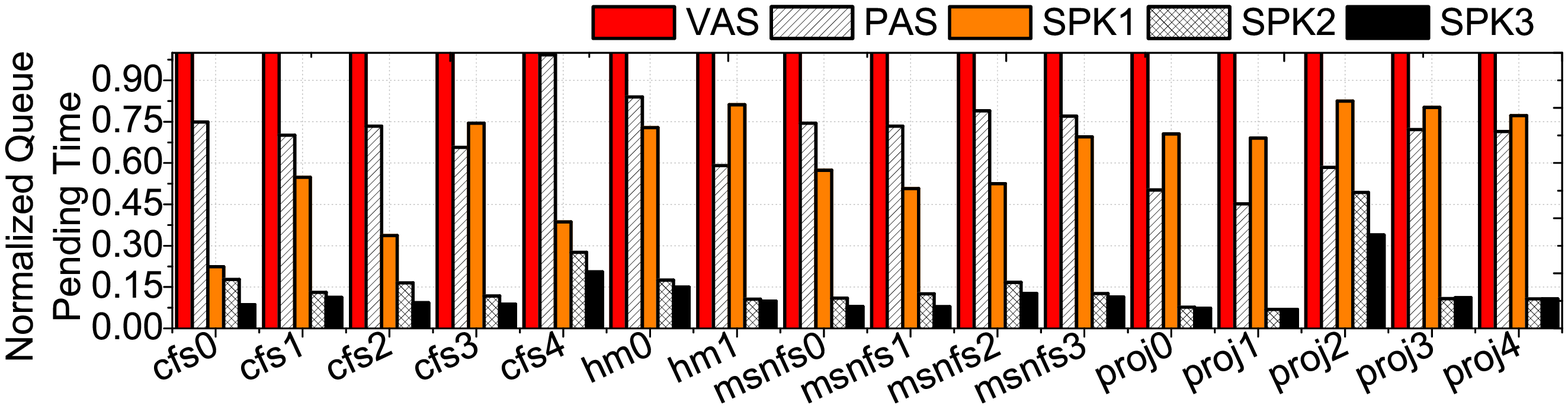}}}


\vspace{-5pt}
\caption{VAS, PAS, SPK1, SPK2, and SPK3 performance comparison.}
\vspace{-10pt}
\label{fig:bargraph}
\end{figure*}

\section{Evaluation}
\label{sec:eval}
 
\subsection{Configuration}
\noindent \textbf{SSD and NAND flash.} We evaluated Sprinkler using a \emph{cycle-accurate} SSD simulator with hardware-validated multiple flash simulation model \cite{nfs} that allows us vary the number of flash chips, ranging from 64 flash chips (8 channels) to 1024 flash chips (32 channels). Each channel works on ONFI 2.x, which is the most popular flash interface specification in the SSD industry. Note that, \emph{considering slow write speed on flash memory array and power/cost of devices, many vendors such as Micron, Fusion-IO and OCZ employ ONFi 2.x, in stead of a 400MHz flash interface, even for high-performance PCIe SSDs} (e,g., Micron P320H, Fusion-IO ioFX series, and OCZ RevoDrive series). For the flash microarchitecture configuration, each flash chip employs two dies and four planes, and each die consists of 8,192 blocks. 128 pages are put together as a single block, and the unit size of each page is 2KB. Our simulation framework can capture the intrinsic write (programming) variation latency \cite{harey, rel:abnormal, nfs, bleak} of Multi Level Cell (MLC) NAND flash memory, varying from 200 $\mu$s (slow page) to 2200 $\mu$s (fast page) \cite{spec:micronmlc} based on the address of the page being accessed. In addition, read latency is configured to be 20 $\mu$s. Note that these values represent a state-of-the-art SSD and NAND flash package. 

\noindent \textbf{Schedulers.} We evaluate five different I/O schedulers:
\begin{itemize}[itemsep=1pt,parsep=1pt,topsep=4pt, partopsep=0pt]
\item VAS	--  Virtual address scheduler, using FIFO. 
\item PAS	-- Physical address scheduler, using extra flash queues.
\item SPK1	-- Sprinkler, using only FARO.
\item SPK2	-- Sprinkler, using only RIOS.
\item SPK3	-- Sprinkler, using both FARO and RIOS.
\end{itemize}
For all the schedulers tested, we introduce a standard command queue, which allows storage devices to execute I/Os in an out-of-order fashion \cite{int:sata}. In this evaluation, PAS is implemented to support system-level, coarse grain, out-of-order execution \cite{o3}, which means that it can skip the busy flash chips and commit the other memory requests to idle chips.

\noindent \textbf{Firmware.} We implemented a pure page-level address mapping FTL and a garbage collection strategy similar to the one employed in \cite{arch:dtssd}. Flash controllers manage flash transactions following the open NAND flash interface specification \cite{int:onfi}. 

\noindent \textbf{Traces.} We employ data center workloads \cite{exp:msr} from  public trace repositories \cite{exp:iotta}. Our traces consist of corporate mail file server (\emph{cfs}), hardware monitor (\emph{hm}), MSN file storage server (\emph{msnfs}), and project directory service (\emph{proj}). The important characteristics of our traces are given in Table \ref{tab:workloads}.   

\begin{figure*}
\centering
\def\subfigcapskip{0pt}
\subfloat[Inter-chip idleness.  ]{\label{fig:inter-chip}\rotatebox{0}{\includegraphics[width=0.5\linewidth]{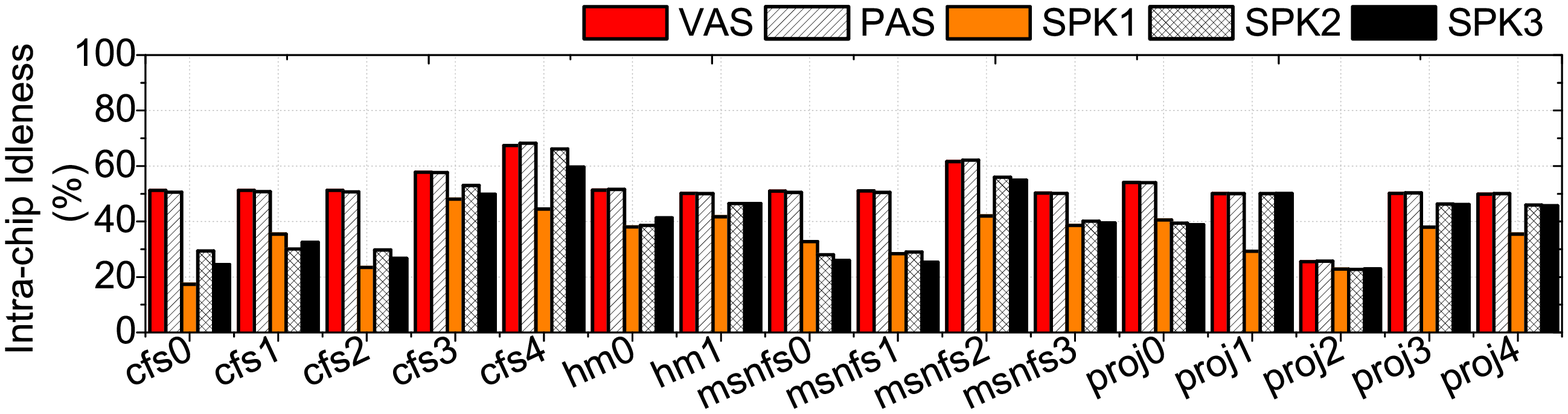}}}
\subfloat[Intra-chip idleness.
]{\label{fig:intra-chip}\rotatebox{0}{\includegraphics[width=0.5\linewidth]{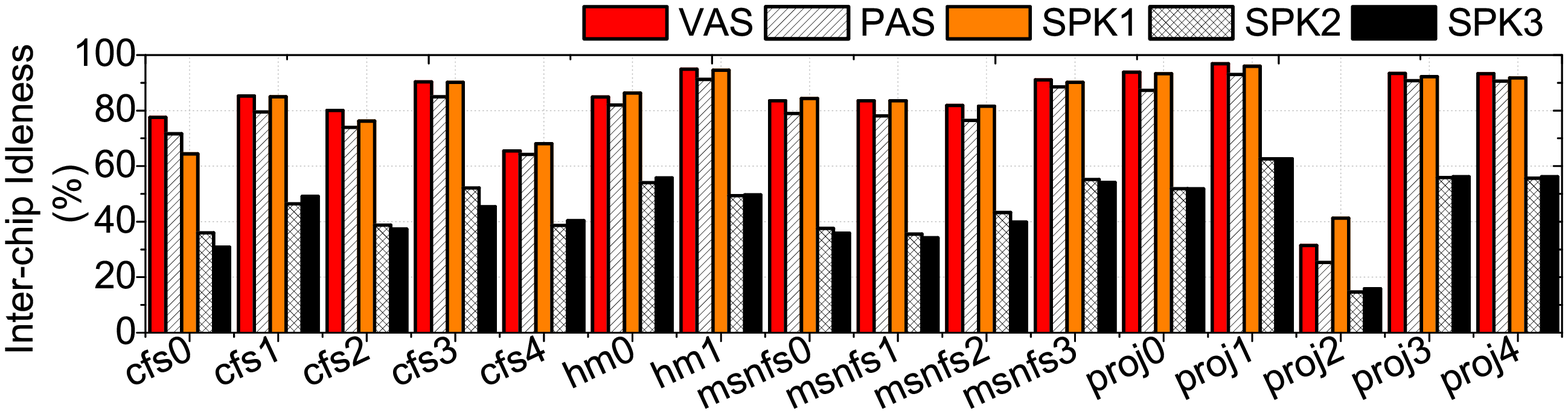}}}
\vspace{-10pt}
\caption{Idleness analysis.}
\vspace{-10pt}
\label{fig:bargraph}
\end{figure*}

\subsection{System Performance}
\noindent \textbf{Bandwidth.} Figure \ref{fig:band} plots the I/O bandwidth for five different I/O schedulers we implemented in NVMHC. One can see from these results that Sprinkler generates better throughput values than VAS and PAS. Specifically, SPK3 boosts the I/O bandwidth by at least 2.2 times, compared to VAS. For all the workloads we tested, the throughput improvement brought by SPK3 over VAS ranges between 42 MB/sec and 300 MB/sec. Further, as compared to PAS, SPK3 provides 1.8 times better throughput. In this case, the improvement ranges between 38MB/sec and  200MB/sec. Our performance improvements are more pronounced when the I/O instruction addresses exhibit higher (potential) transactional-locality (e.g., cfs3, cfs4, msnfs2$\sim$3). 

We also see that SPK3 improves throughput regardless of how read or write intensive a workload is. 
As compared to PAS, SPK1 occasionally hurts performance. This is because, even though FARO is capable of increasing FLP, it cannot always secure enough memory requests to achieve high FLP without RIOS's help. In contrast, SPK2 always outperforms VAS and PAS, and exhibits better performance than SPK1 in most cases. This is because RIOS can build and commit memory requests irrespective of the host level information by employing a fine-grain, out-of-order execution strategy.

\noindent \textbf{IOPS.} Figure \ref{fig:iops} gives the IOPS values achieved by different I/O schedulers. For cfs1, msnfs0$\sim$1 and proj0$\sim$1 workloads whose access patterns include mostly small random requests, the achieved bandwidth improvement is not dramatic, compared to other workloads. In these workloads, SPK2 and SPK3 improve over VAS and PAS by about 2x. 
In contrast, proj2 consists of large I/O requests, which have low transactional-locality. As a result, most schedulers provide low IOPS. Unlike the performance observed under other workloads, SPK1 outperforms SPK2 in this case. Even though the address accesses across requests exhibit low transactional-locality, memory requests that belong to a given I/O request have sequentiality to some extent. As a result, in proj2, SPK2 can improve FLP without any help from RIOS. Further, by combining RIOS and FARO, SPK3 shows better performance than any other scheduler in all workloads. Even with proj2, SPK3 generates about 2x better IOPS than PAS.

\noindent \textbf{Latency and Queue Stall Time.} 
Figures \ref{fig:avg-latency} and \ref{fig:pending} plot average SSD device-level latency and the device-level queue stall time, respectively. In our evaluation, the device-level latency means the response time per I/O request, not per memory request or transaction, and the queue stall time is \emph{normalized} to that of VAS. 
SPK3 successfully reduces the device-level latency from 59.1\% to 92.3\%, as compared to VAS, for all the workloads tested. SPK1 provides worse latency than  PAS under certain workloads such as cfs3, proj0, and proj1, since FARO itself cannot secure enough memory requests and still has parallelism dependency problem with such workloads.  
One can also observe that, all SPK schedulers significantly reduce the queue stall time. 
In particular, the queue stall time experienced with SPK3 is about 86\% less than that of VAS. The shorter queue stall time at the device level increases opportunities for the host level modules to parallelize I/O accesses and reduce the number of blocking I/O requests.

\subsection{Device-Level Idleness}
\noindent \textbf{Inter-chip Idleness.} Figure \ref{fig:inter-chip} shows inter-chip idleness values under different scheduling strategies. Even though PAS and SPK1 take advantage of the physical address space exposure, they cannot reduce inter-chip idleness significantly. The main reason behind this is the fact that they cannot fully parallelize data accesses of incoming I/O requests due to parallelism dependency. In contrast, SPK2 activates as many flash chips as possible by relaxing parallelism dependency, which in turn significantly reduces the inter-chip idleness. Specifically, compared to VAS, SPK3 improves inter-chip idleness by about 46.1\%, on average.

\noindent \textbf{Intra-chip Idleness.}    
Intra-chip idleness (Figure \ref{fig:intra-chip}) paints a different picture, compared to the inter-chip idleness. Even though SPK1 suffers from parallelism dependency, it is in a better position to compose high-FLP transactions than SPK2. Thus, SPK1 reduces intra-chip idleness much more than SPK2. SPK2 is able to reduce intra-chip idleness slightly, although it does not build transactions with high FLP in mind. This is because relaxed parallelism dependency allows flash controllers to take more advantage of die interleaving than PAS or VAS. 
SPK3, employing both FARO and RIOS, performs worse than SPK1 because it introduces, in some cases, more system-level contention across the multiple memory requests composed by RIOS.  It should be noted however that, \emph{when both intra-chip and inter-chip idleness are considered, SPK3 outperforms the remaining schedulers tested}.

\subsection{Time Series Analysis}

\begin{figure}
\centering
\def\subfigcapskip{0pt}
\subfloat[VAS vs. PAS  ]{\label{fig:timeseries-pas}\rotatebox{0}{\includegraphics[width=0.5\linewidth]{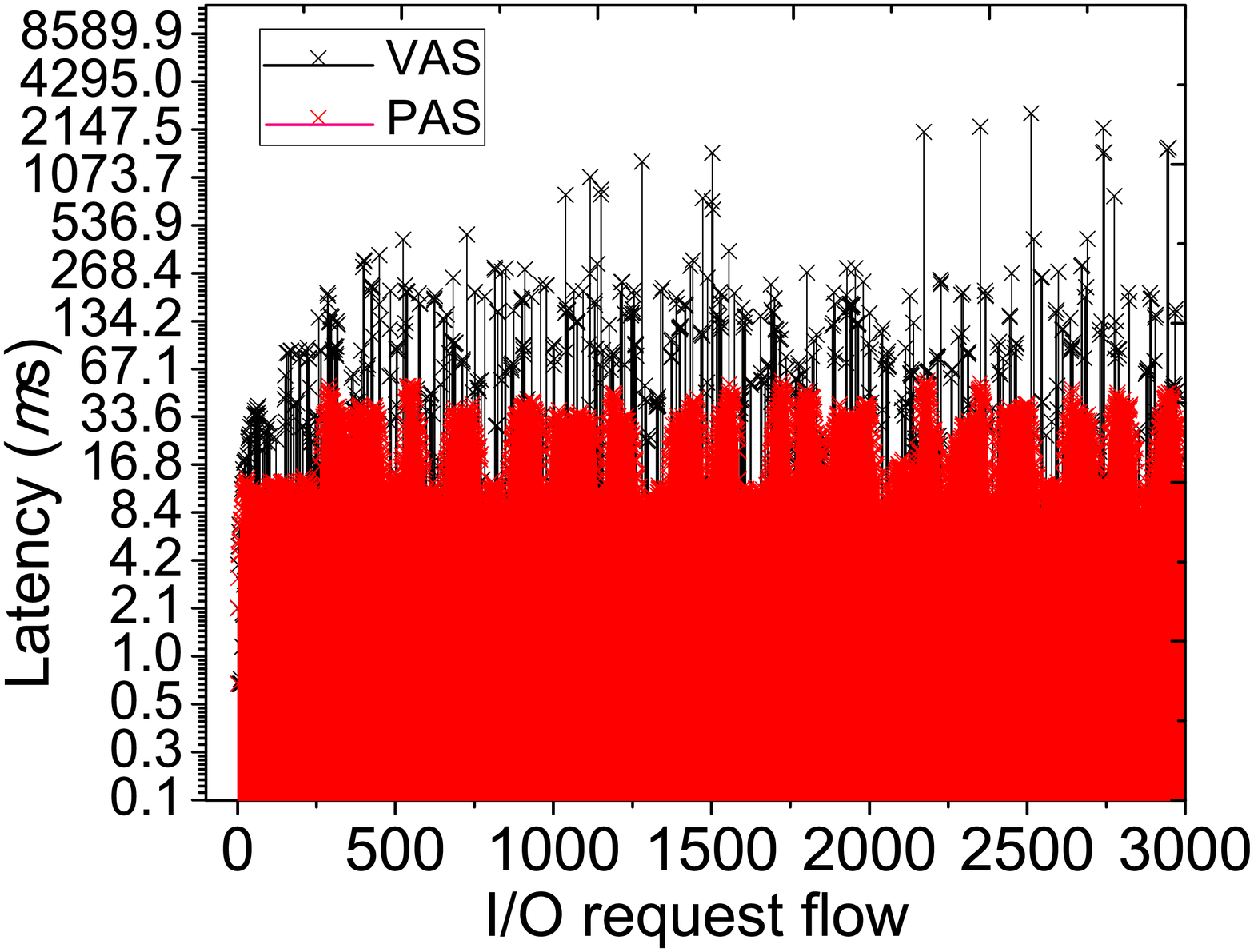}}}
\subfloat[VAS vs. SPK3
]{\label{fig:timeseries-spk3}\rotatebox{0}{\includegraphics[width=0.5\linewidth]{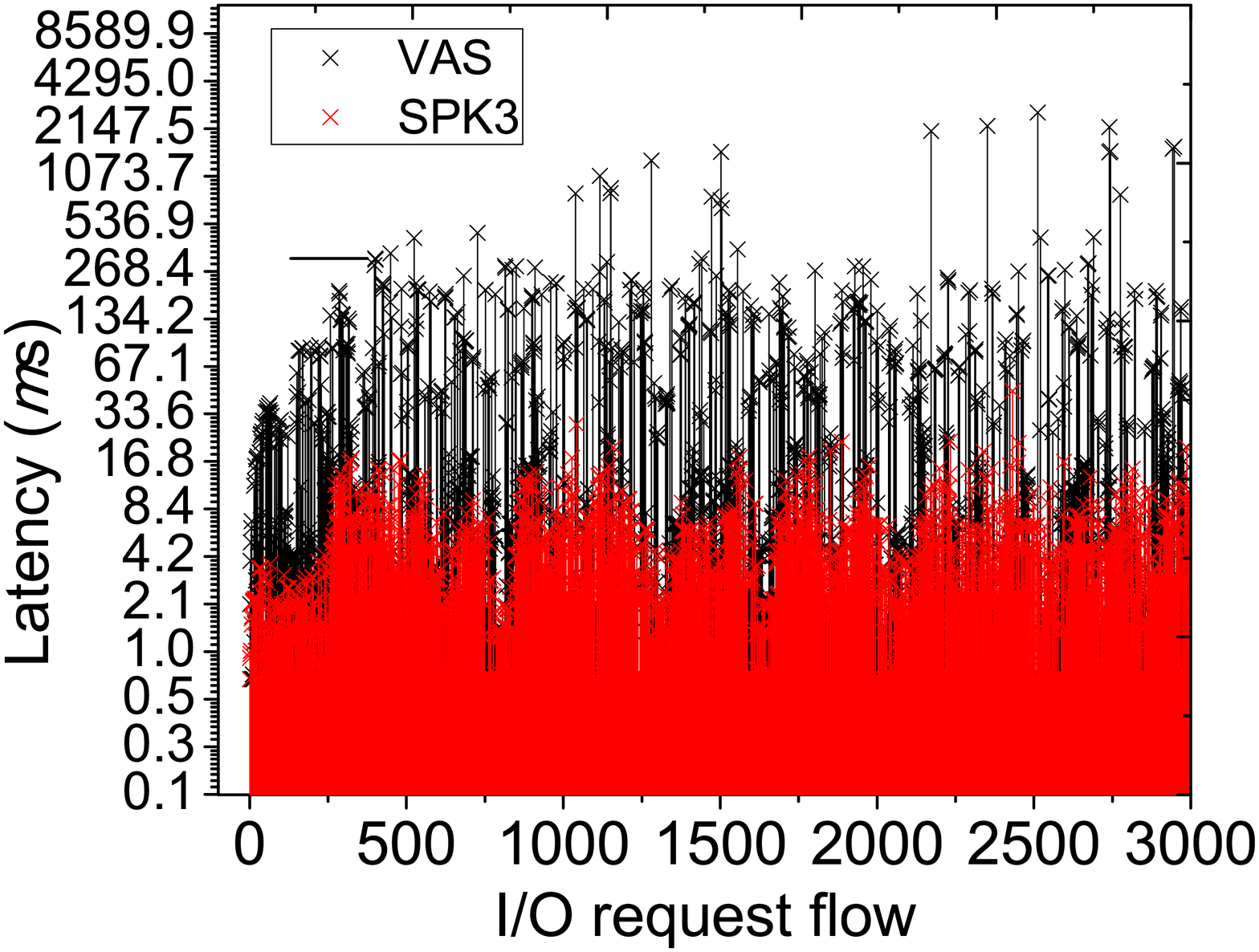}}}
\vspace{-5pt}
\captions{Time series analysis for PAS and Sprinkler (SPK3). SPK3 generates 80\% and 64\% shorter device-level latency than VAS and PAS, respectively.}
\label{fig:timeseries}
\end{figure}

We now compare the device-level latencies experienced by PAS and SPK3 against VAS, using three thousand I/O instructions from the beginning of msnfs1 (one of our traces). Figure \ref{fig:timeseries} clearly shows the superiority of SPK3 over PAS. As shown in Figure \ref{fig:timeseries-pas}, PAS successfully reduces latency of I/O requests, and provides, much more stable performance, compared to VAS.  This is because PAS schedules memory requests using physical addresses, and the extra queues employed for each flash chip allows course-grain out-of-order execution. However, the latency improvement brought by PAS is limited because 1) it does not exploit FLP and 2) the memory request composition and commitment when using PAS are not free from the parallelism dependency problem. 
In contrast, SPK3 provides much shorter latency (which is even better than PAS) since the over-committed memory requests are composed using \emph{fewer} flash transactions, and these memory requests are served \emph{in parallel} by multiple flash chips at the same time, thanks to the relaxed parallelism dependency.  

\subsection{Execution Time Breakdown}

\begin{figure}
\centering
\def\subfigcapskip{0pt}
\subfloat[PAS]{\label{fig:brk-VAS}\rotatebox{0}{\includegraphics[width=0.5\linewidth]{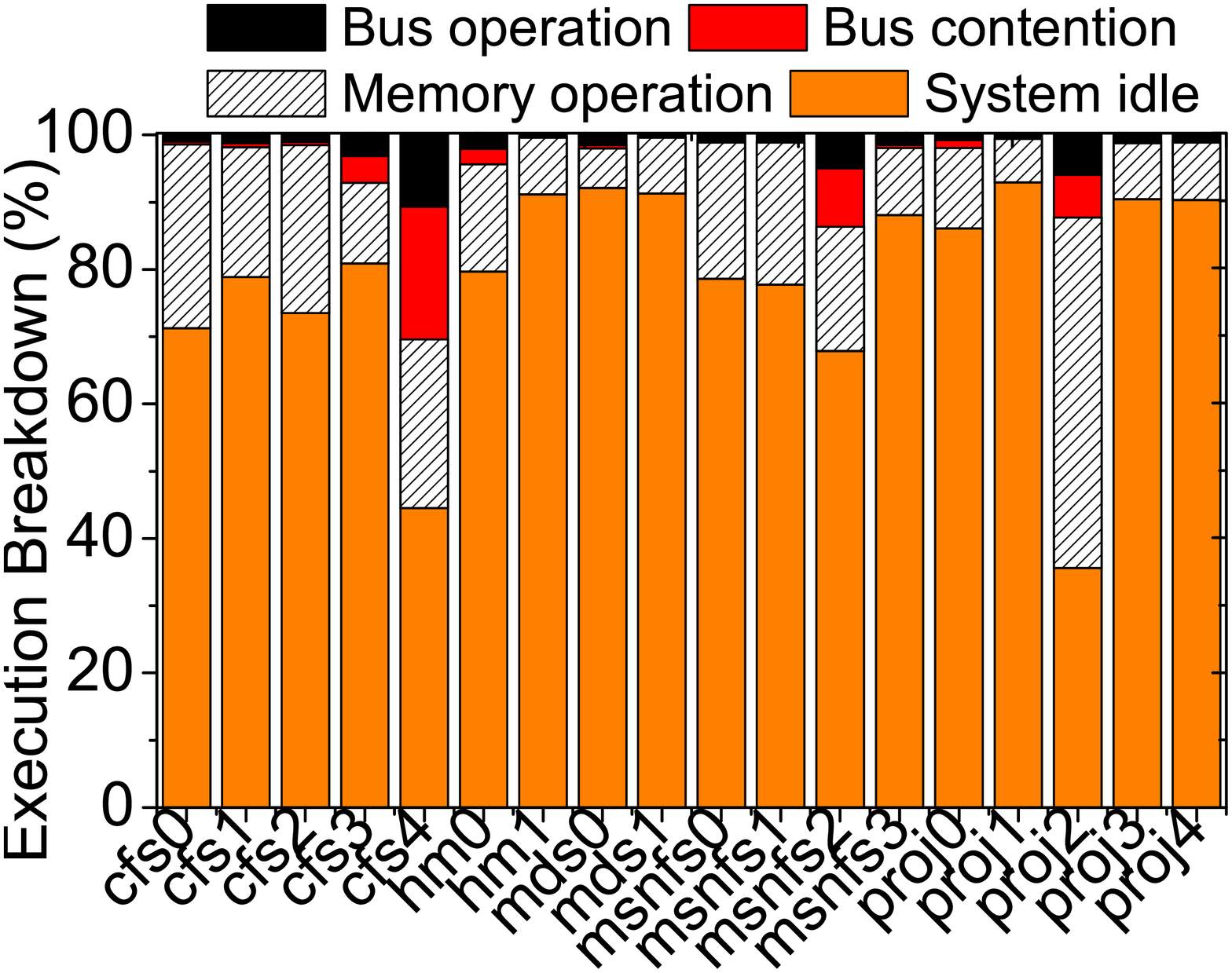}}}
\subfloat[SPK3
]{\label{fig:brk-SPK3}\rotatebox{0}{\includegraphics[width=0.5\linewidth]{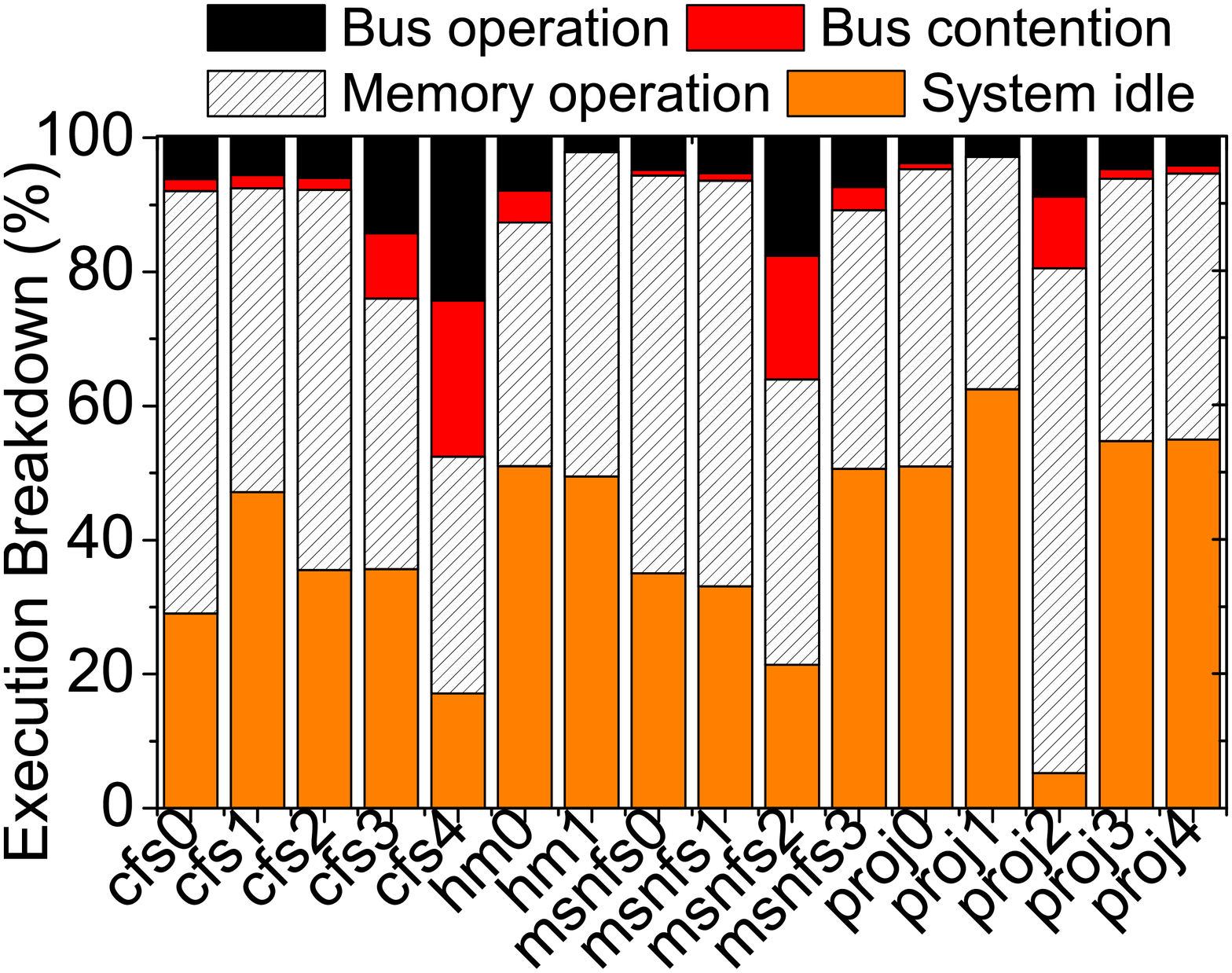}}}
\vspace{-5pt}
\captions{Execution time breakdown. SPK3 eliminates system level idleness by 40.5\% (50.7\%), compared to PAS (VAS).}
\label{fig:exe-brkdown}
\end{figure}

\begin{figure*}
\centering
\def\subfigcapskip{0pt}
\subfloat[PAS  ]{\label{fig:pas-pal-brkdown}\rotatebox{0}{\includegraphics[scale=0.165]{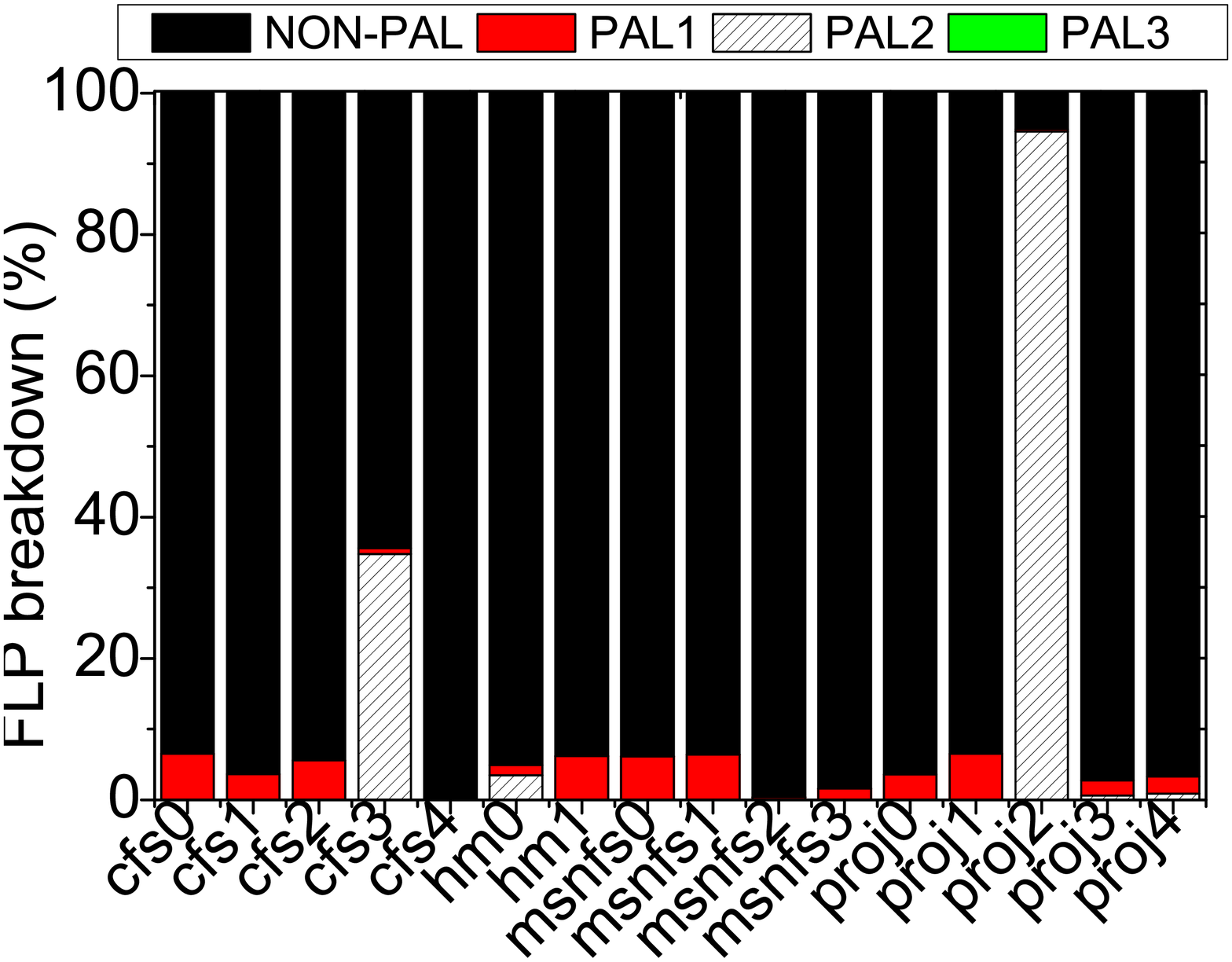}}}
\subfloat[SPK1
]{\label{fig:spk1-pal-brkdown}\rotatebox{0}{\includegraphics[width=0.25\linewidth]{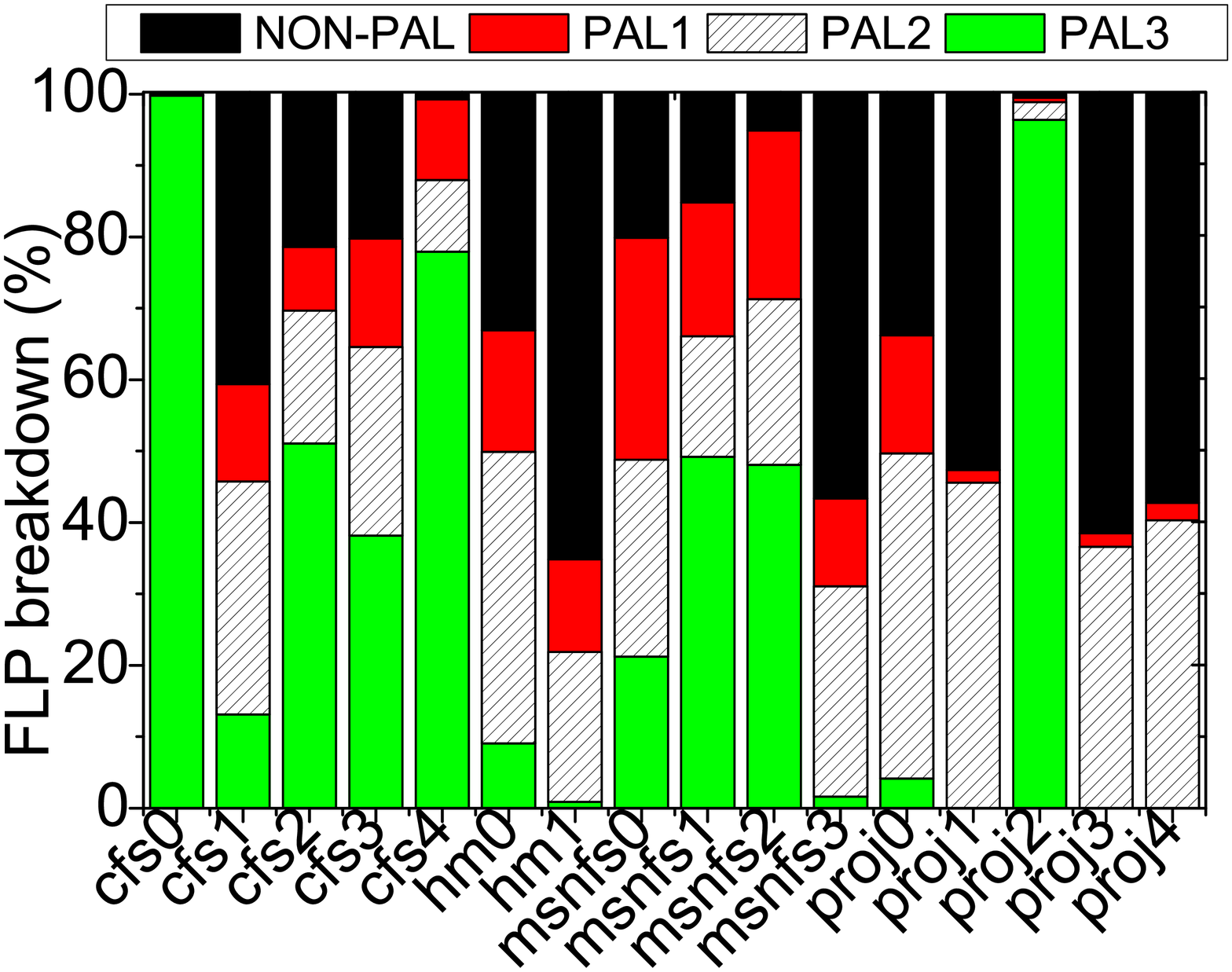}}}
\subfloat[SPK2
]{\label{fig:spk2-pal-brkdown}\rotatebox{0}{\includegraphics[width=0.25\linewidth]{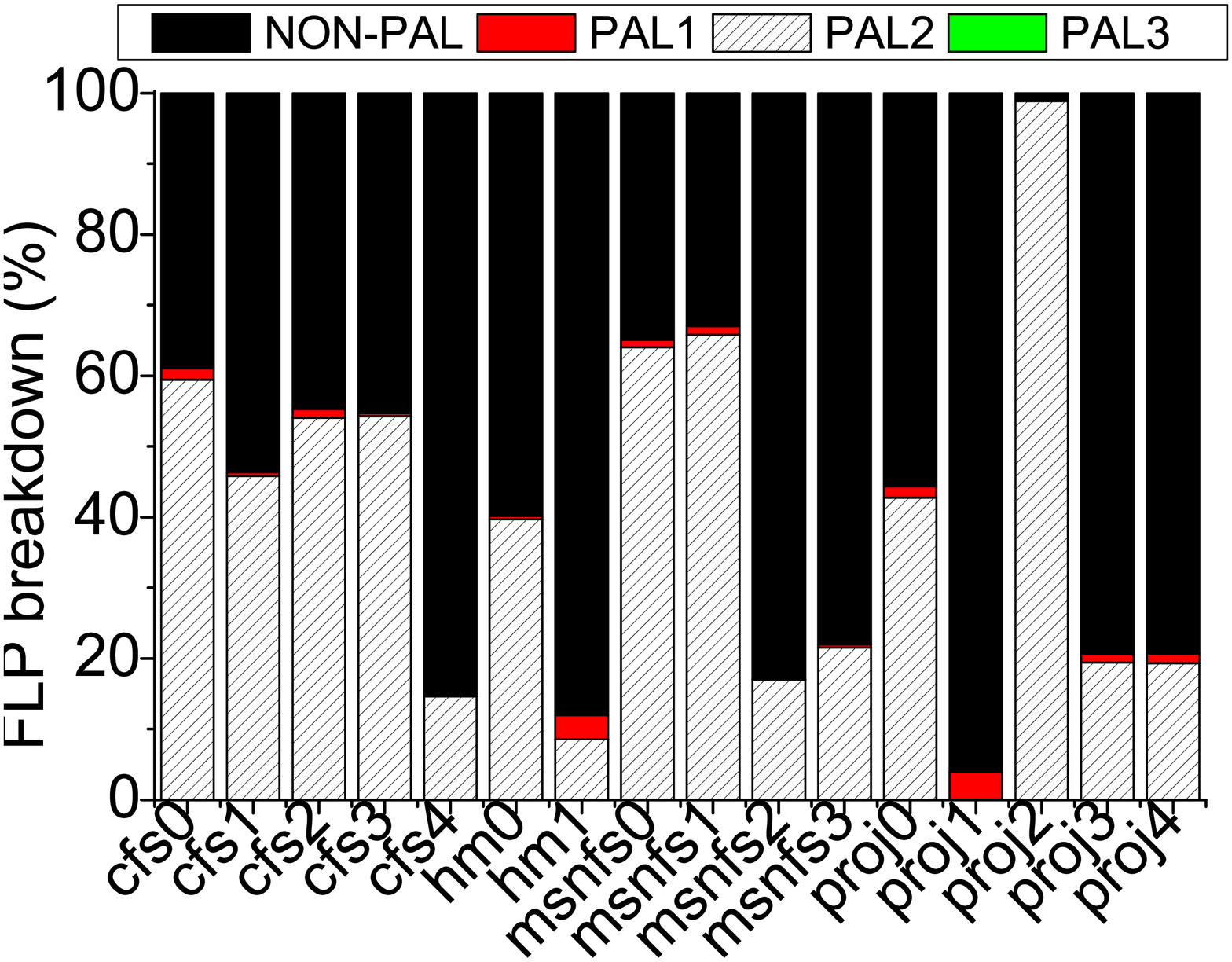}}}
\subfloat[SPK3
]{\label{fig:spk3-pal-brkdown}\rotatebox{0}{\includegraphics[width=0.25\linewidth]{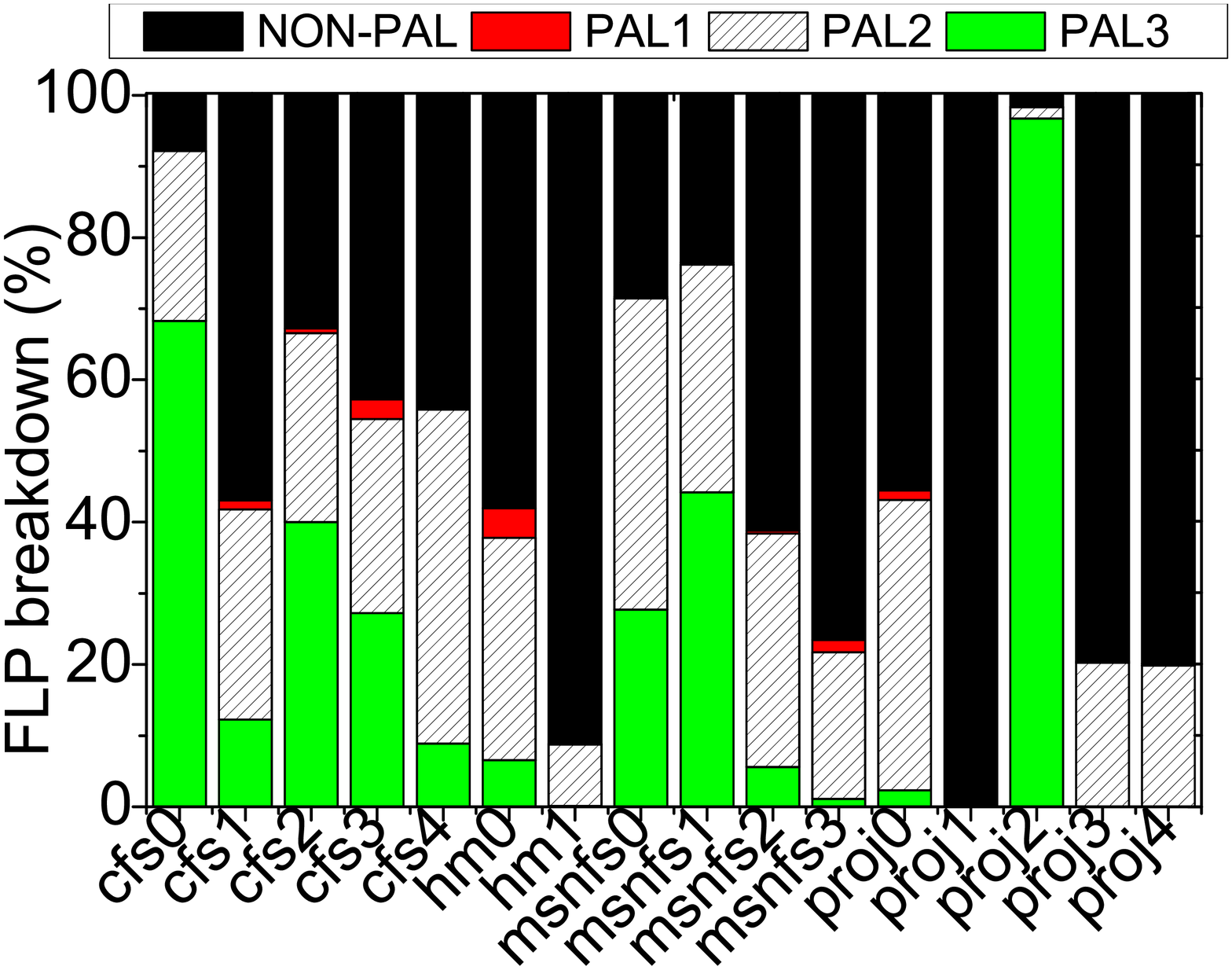}}}
\vspace{-10pt}
\captions{Flash-level parallelism breakdown. While SPK1 (FARO only) maximizes FLP,  SPK3 (FARO+RIOS) carefully balances parallelism between SLP and FLP.}
\label{fig:flp-brkdown}
\end{figure*}

Figure \ref{fig:exe-brkdown} gives a breakdown of the total execution time into bus activate, bus contention, cell activate, and idle time components. As shown in Figure \ref{fig:brk-VAS}, PAS wastes large amounts of time, clearly indicating that it suffers from the low resource utilization problem, as it does not take into account the flash microarchitecture characteristics. In contrast, SPK3 increases the memory cell active time by maximizing FLP as well as relaxing the parallelism dependency, as shown in Figure \ref{fig:brk-SPK3}. Also, the bus contention time increases in SPK3 as a result of increasing the amount of flash memory cell activities in workloads whose fraction of reads is larger than that of writes, such as cfs3, msnfs2, and proj2. However, we still have spare time, which can be utilized to execute I/O instructions, in all the workloads tested.


\subsection{Parallelism Analysis}

Figure \ref{fig:flp-brkdown} decomposes parallelism by four different levels; \emph{NON-PAL} captures the I/O requests that are served by only SLP concurrency oriented strategies, such as channel stripping and pipelining; \emph{PAL1} represents the impact of plane sharing when combined with the SLP concurrency strategies; \emph{PAL2} indicates the impact of die interleaving  combined with the SLP concurrency schemes; and finally, \emph{PAL3} captures the impact when die interleaving and plane sharing are combined with SLP optimization strategies so that I/O requests are fully served with the highest degree of parallelism (4x higher performance than NON-PAL).  While VAS serves I/O requests with only PAL1, contributing to 1\% \textasciitilde 3\% of the total execution, PAS improves parallelism by exploiting the other levels of parallelism, as shown in Figure \ref{fig:pas-pal-brkdown}. However, there is no PAL3 in PAS, and it still experiences low FLP due to the parallelism dependency problem. SPK1 provides the best way of achieving high FLP, but it has a lower system-level chip utilization (see Section \ref{sec:chiputil}). The degree of parallelism in SPK2 is better than that of PAS; however, like PAS, SPK2 does not achieve high levels of FLP.
Lastly, the degree of parallelism obtained by SPK3 is lower than that of SPK1, but it enjoys the benefits of both SPK1 and SPK2 and makes parallelism more balanced between SLP and FLP, thereby achieving high levels of parallelism as well as high chip utilization.

\begin{figure*}
\centering
\def\subfigcapskip{0pt}
\subfloat[64 flash chips]{\label{fig:ch-8-flash-util}\rotatebox{0}{\includegraphics[scale=0.2]{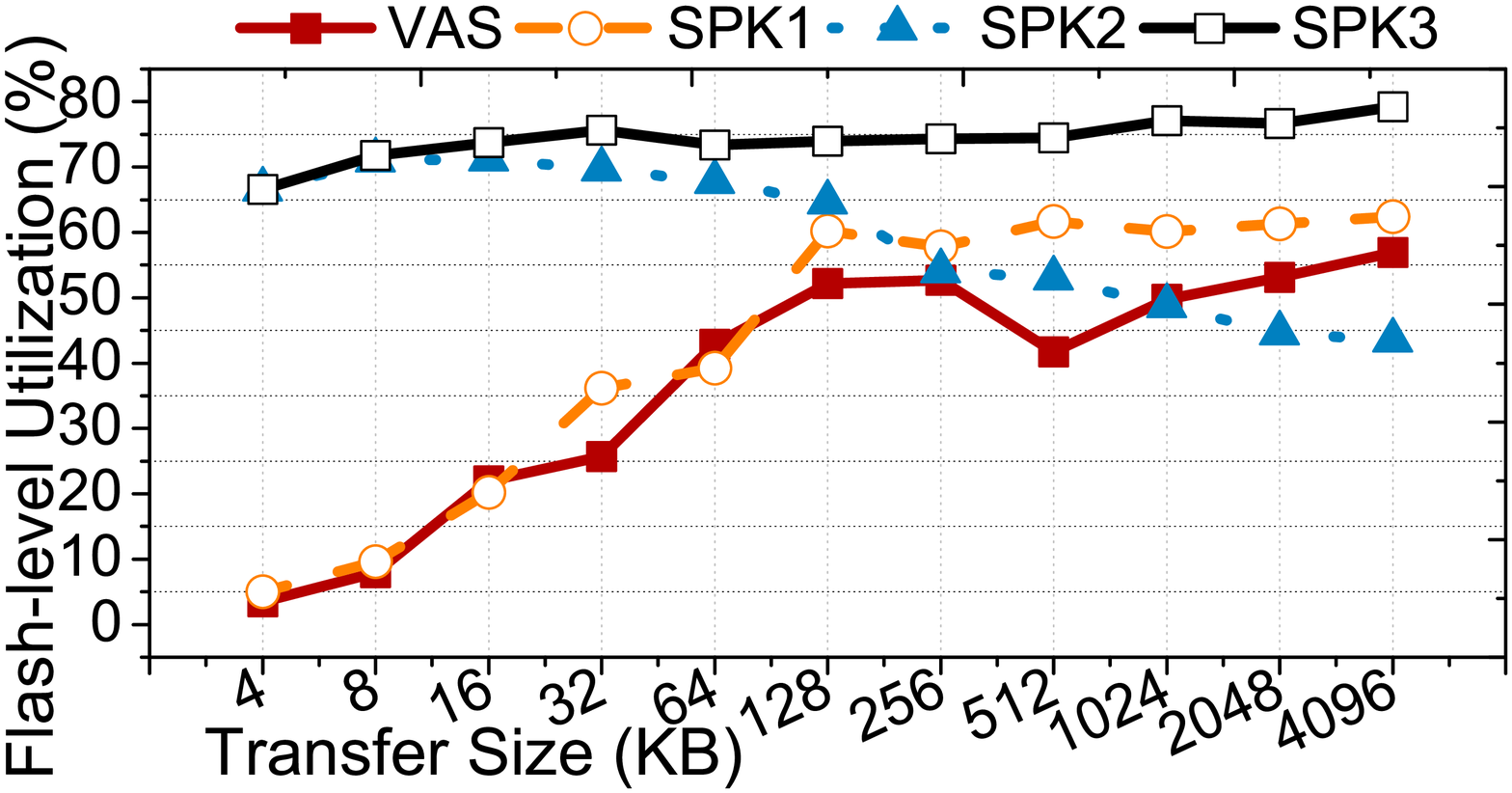}}}
\quad
\subfloat[256 flash chips
]{\label{fig:ch-16-flash-util}\rotatebox{0}{\includegraphics[scale=0.2]{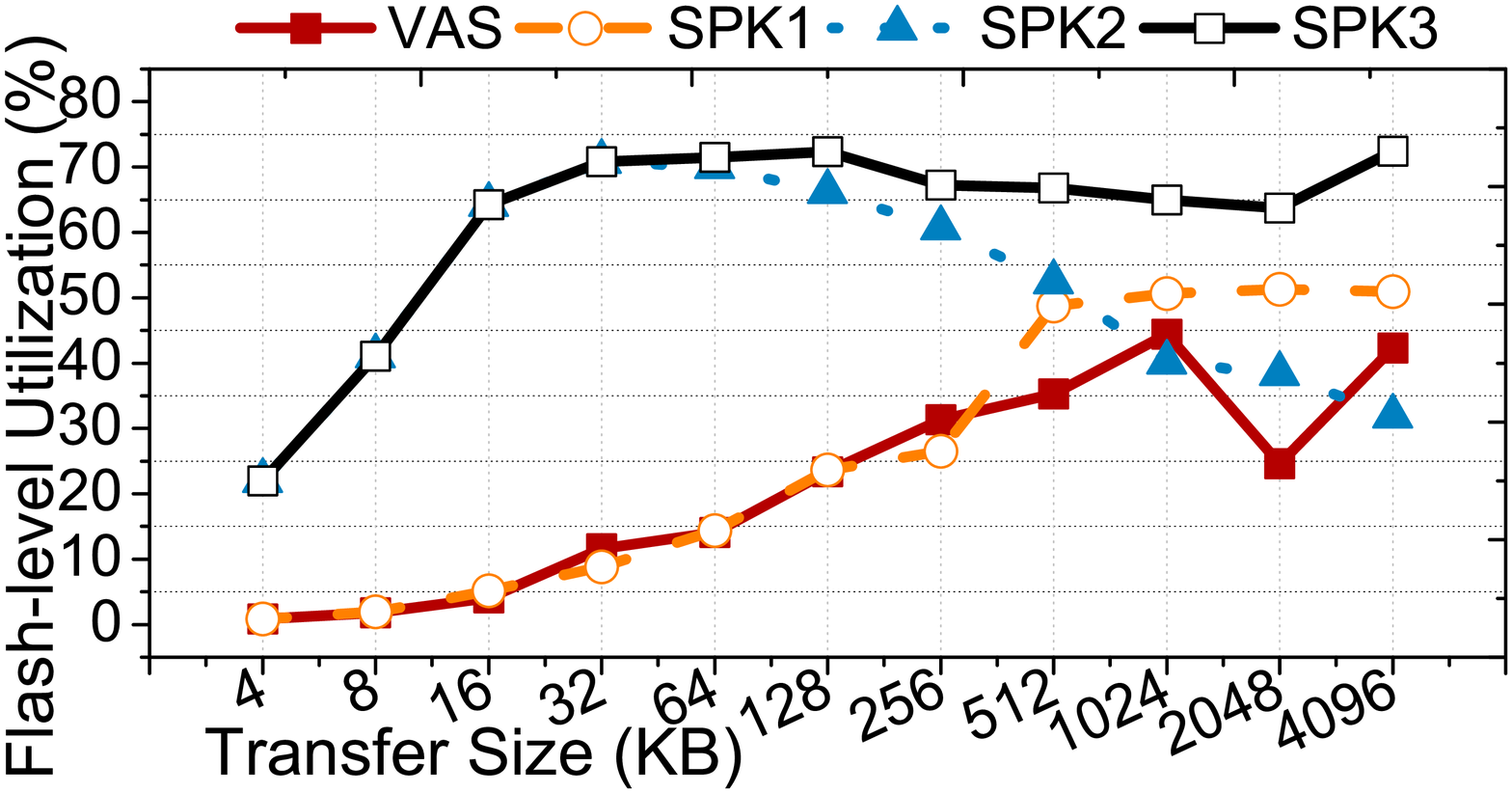}}}
\quad
\subfloat[1024 flash chips
]{\label{fig:ch-32-flash-util}\rotatebox{0}{\includegraphics[scale=0.2]{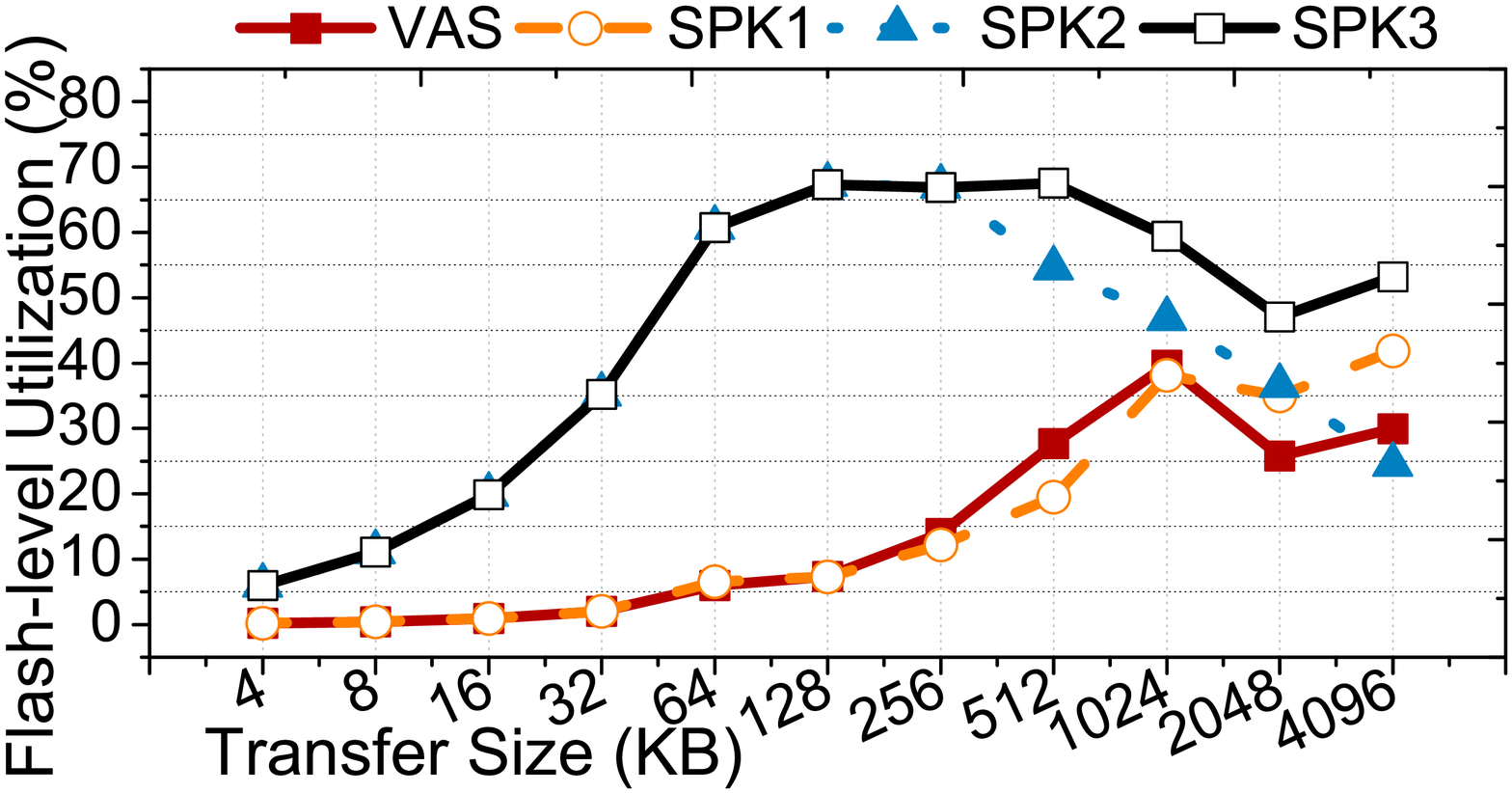}}}
\vspace{-10pt}
\captions{Chip utilization analysis. SPK3 outperforms other schedulers, irrespective of data transfer size and SSD internal configuration.}
\label{fig:util-eval}
\end{figure*}

\subsection{Resource Utilization Analysis}

\label{sec:chiputil}
Figure \ref{fig:util-eval} plots the chip utilization results when varying the transfer sizes from 4KB to 4MB, and varying the number of flash chips from 64 to 1024. In general, the chip utilization of VAS keeps increasing as the transfer size increases. However, in some cases where the length of the request (the number of memory requests associated an I/O) spans all chips, but not all the flash internals (512KB, 1MB, 2MB in 64 chips, 256 chips and 128 chips, respectively), the utilization of VAS drops to some extent because it simply strips an I/O request across multiple resources in a round-robin fashion without taking into account the underlying flash microarchitecture. 
However, as the data size keeps continuing to increase, the chip utilization increases again. This is because a larger I/O size covers more flash internals, and this hleps to improve FLP. 

Unlike VAS, our SPKs exhibit different utilization characteristics.  
As shown in the figure, SPK1 can improve the chip utilization by 16\% only if incoming I/O request sizes are large. This is because in this case it can secure enough memory requests that belong to a single I/O request to compose a high-FLP transaction. However, it does not work well when request sizes are small, due to strong parallelism dependency.   
In contrast, SPK2 shows better chip utilization only when the data sizes of the incoming requests are small. Even tough SPK2 can capitalize on internal resource concurrency by relaxing parallelism dependency, it suffers from the system-level resource contention.
Lastly, \emph{SPK3 shows excellent and sustainable chip utilization}. In this case, FARO consumes as many memory requests as possible which are generated by RIOS. Specifically, overall chip utilizations with 64, 256, and 1024 chips are 71.2\%, 61.5\%, and 44.9\%, respectively, while the corresponding utilization values for VAS are 37\%, 21.2\%, and 13.9\%, in that order.


\begin{figure}
\centering
\def\subfigcapskip{0pt}
\subfloat[64 flash chips  ]{\label{fig:ch-8-trans}\rotatebox{0}{\includegraphics[width=0.5\linewidth]{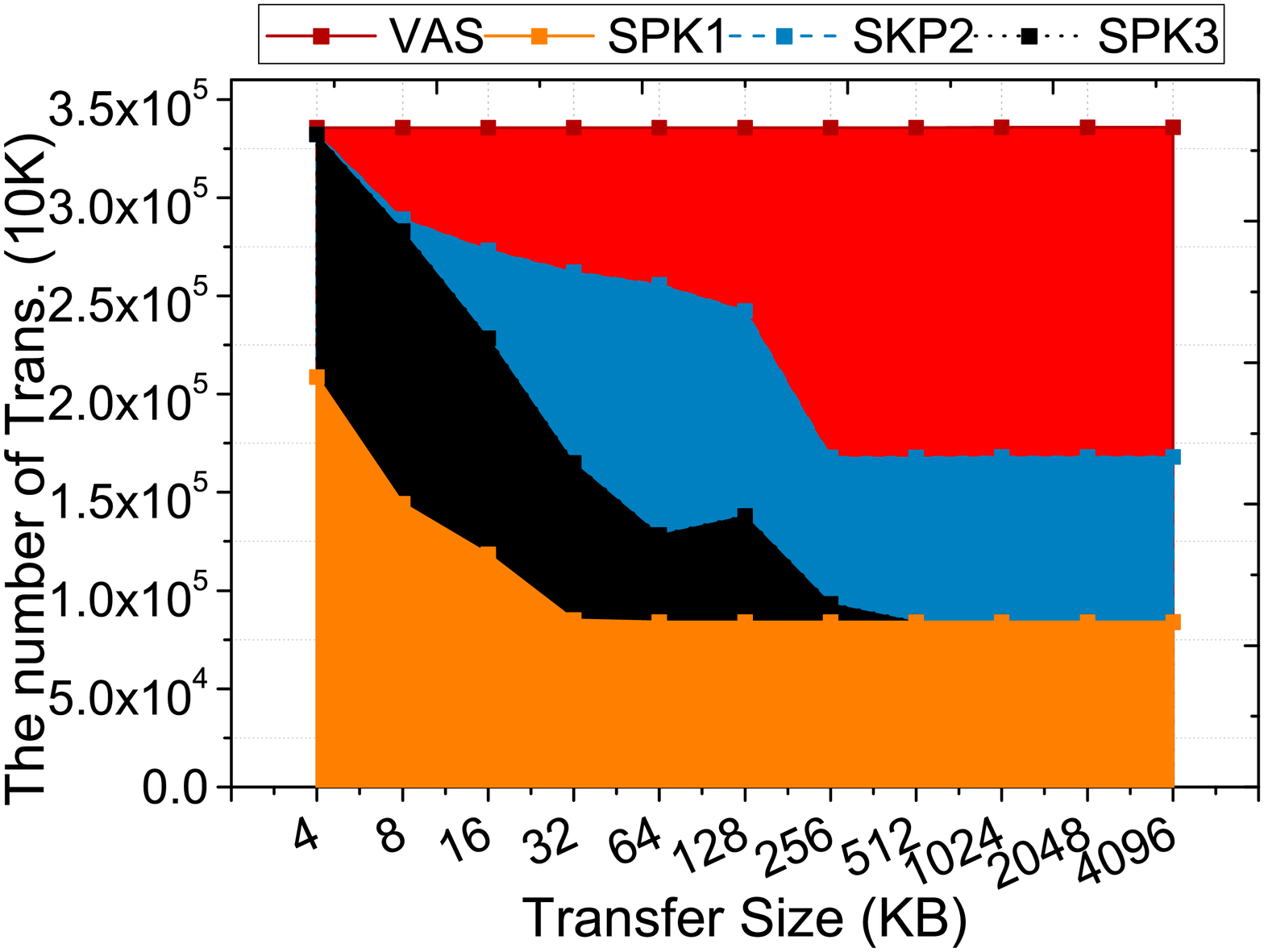}}}
\subfloat[1024 flash chips
]{\label{fig:ch-32-trans}\rotatebox{0}{\includegraphics[width=0.5\linewidth]{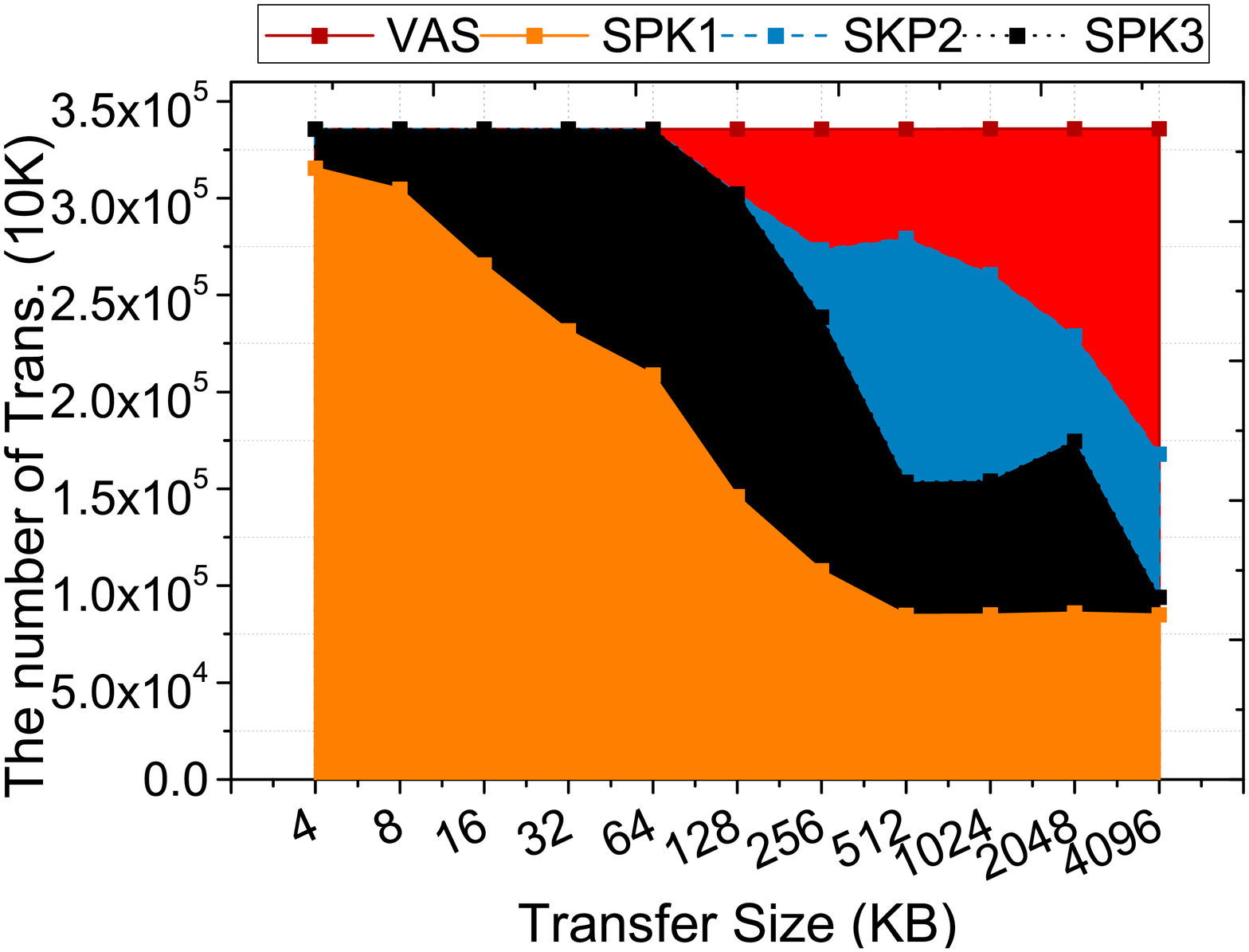}}}
\vspace{-5pt}
\captions{Flash transaction reduction rates. SPK3 reduces the number flash transactions by about 50.2\% than VAS.}
\label{fig:trans-reduct}
\end{figure}

\subsection{Memory Transaction Reduction Rate}
As shown in Figure \ref{fig:trans-reduct}, the over-commitment strategy employed by FARO has a great impact on reducing the number of transactions at a chip level. In contrast, the transaction reduction success of SPK2 is not very high, and gets even worse as the number of flash chips increases. This is because SPK2 parallelizes data access among system-level resources, which leads to low transactional-locality.  Therefore, the flash controller could not secure enough memory requests to coalesce them into a single transaction. SPK3 generates better data reduction rate (50.2\% on average) than SPK2 and enjoys better SLP than SPK1 by employing both FARO and RIOS. 


\subsection{Migration and Re-addressing Callback Impacts}
\begin{figure}
\centering
\def\subfigcapskip{0pt}
\subfloat[64 flash chips  ]{\label{fig:ch8-gc}\rotatebox{0}{\includegraphics[scale=0.165]{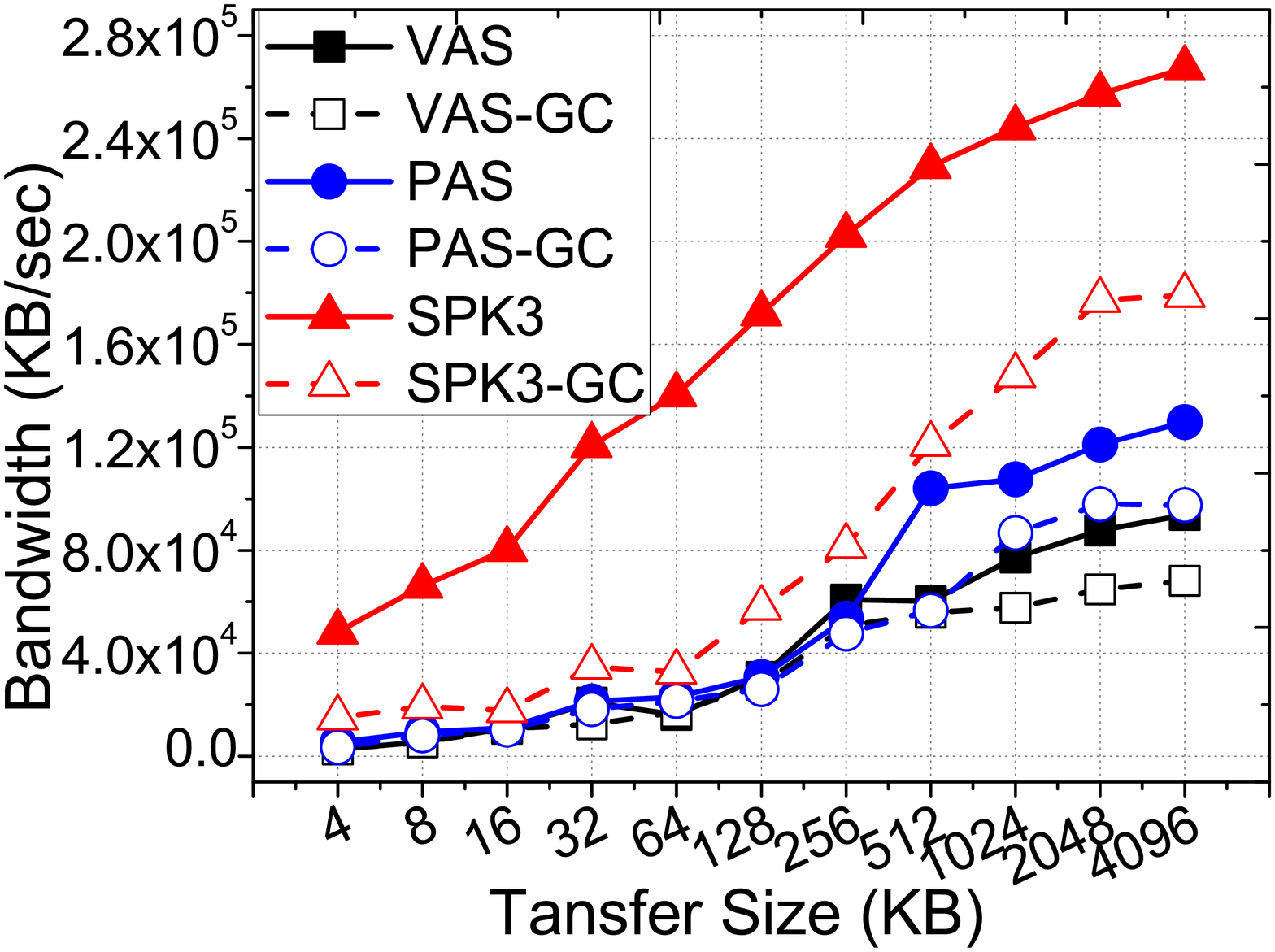}}}
\subfloat[256 flash chips
]{\label{fig:ch16-gc}\rotatebox{0}{\includegraphics[scale=0.165]{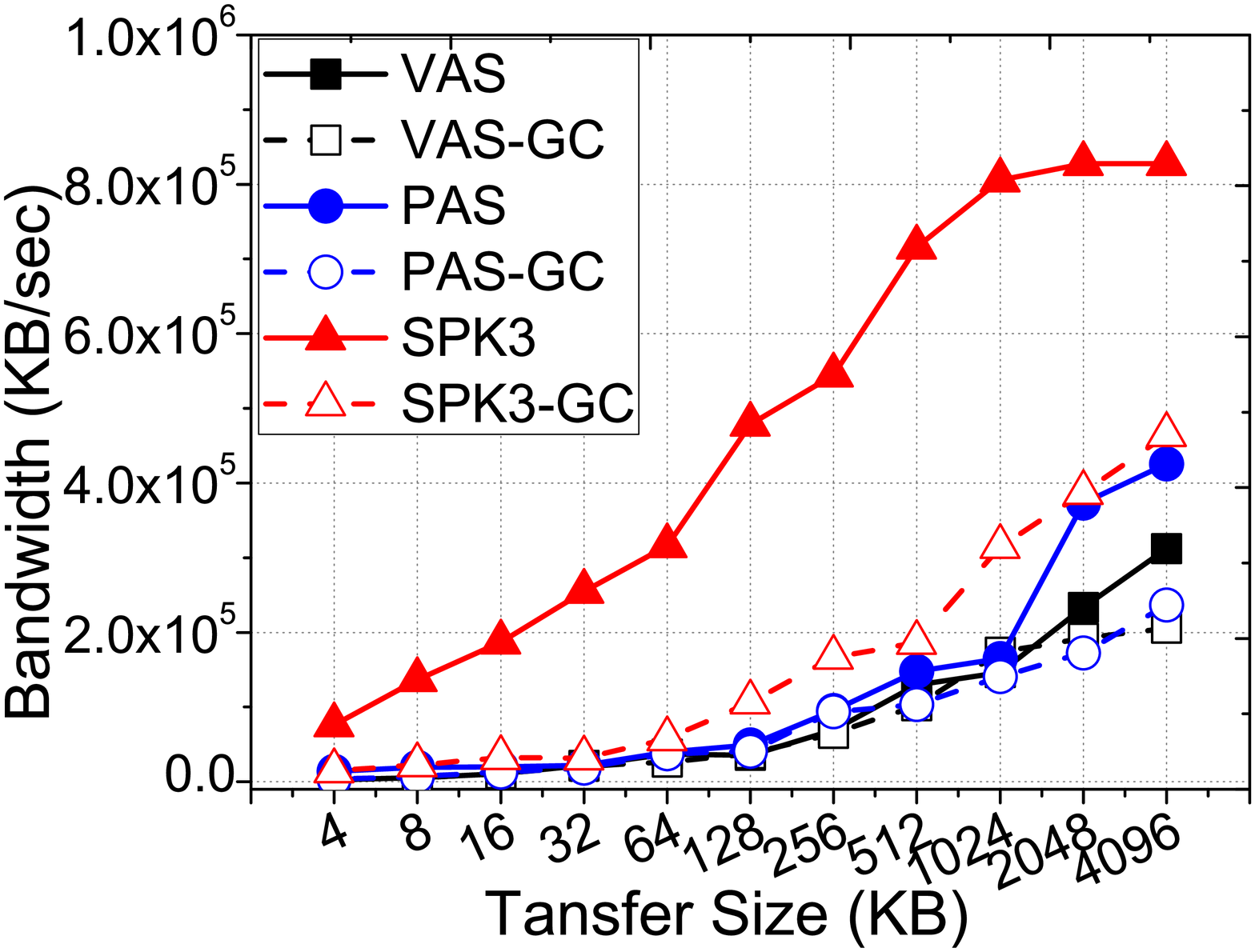}}}
\vspace{-5pt}
\caption{Garbage collection and readdressing impact. SPK3 generates about 2x\% better performance than VAS (and PAS as well) by efficiently updating the physical layout via readdressing callback.}
\label{fig:gc-impact}
\vspace{-5pt}
\end{figure}

Since garbage collection (GC) is one of the most time consuming tasks, and is one of the most frequently-occurring activities during live data migration, we next stressed VAS, PAS, and SPK3 by artificially introducing a very high number of GCs. In this experiment, VAS and PAS have no readdressing callback. We prepared pristine-state SSDs for non-GC evaluation and fragmented SSDs for GC evaluation, which were filled by 95\% with 1 MB random writes (just before the GC begins). As shown in Figure \ref{fig:gc-impact}, all the schedulers tested suffer from performance degradation once the underlying FTL starts performing garbage collections. 

SPK3 exhibits 33\% $\sim$ 78\% performance degradation while VAS experiences 11 $\sim$ 28\% performance degradation. The main reason behind this performance degradation of SPK3 is that many memory requests are stalled at the system level due to the extra read and write activities, caused by GCs. In other words, relaxing parallelism dependency and achieving high transaction-locality are difficult because of GCs. However, SPK3's performance with GCs is still much better than PAS and VAS because it successfully secures new information through readdressing callback after following the GC.  As a result, SPK is able to spread the remaining memory requests over multiple resources and coalesce them again.

\section{Related Work and Discussion}

\noindent \textbf{DRAM controller.} Balancing timing constraints, fairness and different dimension of physical parallelism has long been a problem addressed by DRAM based memory controllers \cite{tcm,atlas,rafique,stfq,davidson,rambus,zuravleff,hur,zuravleff}. SSDs however have sharp differences in device characteristics, form factor (which leads diverse architecture configurations), asymmetric latencies, diverse commands and interface protocols. Consequently, these differences lead to separate research directions and needs for request scheduling and memory controller design.   

\noindent \textbf{Parallelism.}	Prior studies recognize the need to exploit parallelism in flash-based SSDs \cite{hpc:gordon, chen} and propose concurrency-centric methods \cite{dirik, arch:dtssd, kang:multichannel}. For exploiting different levels of internal parallelism, different page allocation strategies \cite{shin, hu, pas} have also been  investigated. 	Even though all these studies demonstrate significant performance improvements and better parallelism over more serial alternatives, concurrency methods and page allocation strategies are typically fixed at the SSD design time, and thus they are not in a position to take advantage of parallelism through I/O request scheduling. 

\noindent \textbf{Request scheduling.} In \cite{park, pas}, the authors attempt to uncover the specific resource contention that occurs in SSDs, the areas where parallelism is far below optimal, and present a dynamic request rescheduling scheme to improve performance. However, they fail to account for the reality that the addresses that correspond to requests that they ``reschedule'' are virtual.

\noindent \textbf{Out-of-oder execution.} Ozone \cite{o3} and PAQ \cite{paq} dynamically schedule I/O requests based on physical addresses. Ozone does not wait for the request that create conflicts at the system level, and serves other request in an out-of-oder fashion. However, \emph{Ozone does not consider flash microarchitecture and corresponding FLP}. Further, it requires a hardware-assisted preprocessor, a postprocessor, extra queues, and a reservation station to exploit physical address. In contrast, PAQ is a software-driven dynamic scheduler, which avoids resource conflicts and improves parallelism. PAQ  executes multiple transactions in an out-of-order fashion by being aware of SLP and plane-level parallelism. Note that, even though \emph{PAQ considers resource conflicts, it is limited to handle only read requests}. 

\noindent \textbf{Native command queue (NCQ).} Recall that, unlike DRAM memory requests, an SSD I/O request consists of multiple page-level memory requests whose size varies based on host-side application characteristics. Consequently, even though a device-level queue (such as native command queue and tagged command queue) allows VAS/PAS to reorder incoming \emph{I/O requests} in an out-of-oder fashion, it is difficult to construct a transaction with high FLP by coalescing memory requests that are scattered across multiple I/O requests.

At a high level, all the studies mentioned in this section ignore internal resource utilization and architectural challenges exhibited by state-of-the-art many-chip SSDs. 

\section{Conclusions}
In this paper, we propose  Sprinkler, a novel device-level SSD controller, which targets  maximizing resource utilization and achieving high performance. Specifically, Sprinkler relaxes parallelism dependency by scheduling I/O requests based on internal resource layout, instead of the order imposed by the device-level queue. 
Our extensive experimental evaluation using a cycle-accurate SSD simulation model shows that a many-chip SSD equipped with Sprinkler provides at least 56.6\% shorter latency and 1.8 $\sim$ 2.2 times better throughput than the modern SSD controllers. 

\nocite{ex1,ex2}
\bibliographystyle{latex8}
\bibliography{sprinkler}

\begin{thebibliography}{10}\setlength{\itemsep}{-1ex}\small

\bibitem{arch:dtssd}
N.~Agrawal, V.~Prabhakaran, T.~Wobber, J.~D. Davis, M.~Manasse, and
  R.~Panigrahy.
\newblock Design tradeoffs for {SSD} performance.
\newblock In {\em {USENIX ATC}}, 2008.

\bibitem{hpc:iointensive}
A.~M. Caulfield, J.~Coburn, T.~Mollov, A.~De, A.~Akel, J.~He, A.~Jagatheesan,
  R.~K. Gupta, A.~Snavely, and S.~Swanson.
\newblock Understanding the impact of emerging non-volatile memories on
  high-performance, {IO}-intensive computing.
\newblock In {\em SC}, 2010.

\bibitem{moneta}
A.~M. Caulfield, A.~De, J.~Coburn, T.~I. Mollov, R.~K. Gupta, and S.~Swanson.
\newblock Moneta: A high-performance storage array architecture for
  next-generation, non-volatile memories.
\newblock In {\em MICRO}, 2010.

\bibitem{hpc:gordon}
A.~M. Caulfield, L.~M. Grupp, and S.~Swanson.
\newblock Gordon: Using flash memory to build fast, power-efficient clusters
  for data-intensive applications.
\newblock In {\em ASPLOS}, 2009.

\bibitem{chen}
F.~Chen, R.~Lee, and X.~Zhang.
\newblock Essential roles of exploiting internal parallelism of flash memory
  based solid state drives in high-speed data processing.
\newblock In {\em HPCA}, 2011.

\bibitem{davidson}
J.~W. Davidson and S.~Jinturkar.
\newblock Memory access coalescing: a technique for eliminating redundant
  memory accesses.
\newblock {\em SIGPLAN Not.}, 1994.

\bibitem{dirik}
C.~Dirik and B.~Jacob.
\newblock The performance of {PC} solid-state disks ({SSDs}) as a function of
  bandwidth, concurrency, device architecture, and system organization.
\newblock In {\em ISCA}, 2009.

\bibitem{rel:abnormal}
L.~M. Grupp, A.~M. Caulfield, J.~Coburn, S.~Swanson, E.~Yaakobi, P.~H. Siegel,
  and J.~K. Wolf.
\newblock Characterizing flash memory: Anomalies, observations,and
  applications.
\newblock In {\em MICRO}, 2009.

\bibitem{bleak}
L.~M. Grupp, J.~D. Davis, and S.~Swanson.
\newblock The bleak future of nand flash memory.
\newblock In {\em {FAST}}, 2012.

\bibitem{harey}
L.~M. Grupp, J.~D. Davis, and S.~Swanson.
\newblock The harey tortoise: Managing heterogeneous write performance in
  {SSD}s.
\newblock In {\em {USENIX ATC}}, 2013.

\bibitem{rambus}
S.~I. Hong, S.~A. McKee, M.~H. Salinas, R.~H. Klenke, J.~H. Aylor, and W.~A.
  Wulf.
\newblock Access order and effective bandwidth for streams on a direct rambus
  memory.
\newblock {\em HPCA}, 1999.

\bibitem{amp}
X.-Y. Hu, E.~Eleftheriou, R.~Haas, I.~Iliadis, and R.~Pletka.
\newblock Write amplification analysis in flash-based solid state drives.
\newblock In {\em SYSTOR}, 2009.

\bibitem{hu}
Y.~Hu, H.~Jiang, D.~Feng, L.~Tian, H.~Luo, and S.~Zhang.
\newblock Performance impact and interplay of {SSD} parallelism through
  advanced commands, allocation strategy and data granularity.
\newblock In {\em ISC}, 2011.

\bibitem{spec:nvme}
A.~Huffman.
\newblock {NVM} express 1.0.
\newblock 2011.

\bibitem{hur}
I.~Hur and C.~Lin.
\newblock Adaptive history-based memory schedulers for modern processors.
\newblock 2006.

\bibitem{pas}
M.~Jung and M.~Kandemir.
\newblock An evaluation of different page allocation strategies on high-speed
  {SSD}s.
\newblock In {\em HotStorage}, 2012.

\bibitem{darksecret}
M.~Jung and M.~Kandemir.
\newblock Revisiting widely held ssd expectations and rethinking system-level
  implications.
\newblock In {\em {SIGMETRICS}}, 2013.

\bibitem{agcdgc}
M.~Jung, R.~Prabhakar, and M.~T. Kandemir.
\newblock Taking garbage collection overheads off the critical path in {SSDs}.
\newblock In {\em Middleware}, 2012.

\bibitem{nfs}
M.~Jung, E.~Wilson, D.~Donofrio, J.~Shalf, and M.~Kandemir.
\newblock {NANDFlashSim}: Intrinsic latency variation aware {NAND} flash memory
  system modeling and simulation at microarchitecture level.
\newblock In {\em MSST}, 2012.

\bibitem{paq}
M.~Jung, E.~H. Wilson, III, and M.~Kandemir.
\newblock Physically addressed queueing ({PAQ}): Improving parallelism in solid
  state disks.
\newblock In {\em ISCA}, 2012.

\bibitem{kang:multichannel}
J.-U. Kang, J.-S. Kim, C.~Park, H.~Park, and J.~Lee.
\newblock A multi-channel architecture for high performance {NAND} flash-based
  storage system.
\newblock {\em JSA}, 2007.

\bibitem{sched:ssdaware}
J.~Kim, Y.~Oh, E.~Kim, J.~Choi, D.~Lee, and S.~H. Noh.
\newblock Disk schedulers for solid state drives.
\newblock In {\em EMSOFT}, 2009.

\bibitem{atlas}
Y.~Kim, D.~Han, O.~Mutlu, and M.~Harchol-Balter.
\newblock Atlas: A scalable and high-performance scheduling algorithm for
  multiple memory controllers.
\newblock In {\em HPCA}, 2010.

\bibitem{tcm}
Y.~Kim, M.~Papamichael, O.~Mutlu, and M.~Harchol-Balter.
\newblock Thread cluster memory scheduling: Exploiting differences in memory
  access behavior.
\newblock In {\em MICRO}, 2010.

\bibitem{spec:micronmlc}
{Micron, Inc}.
\newblock {NAND} flash {MLC} datasheet, {MT29F8G08MAAWC}, {MT29F16G08QASWC}.
\newblock In {\em http://www.micron.com/}.

\bibitem{stfq}
O.~Mutlu and T.~Moscibroda.
\newblock Stall-time fair memory access scheduling for chip multiprocessors.
\newblock In {\em MICRO}, 2007.

\bibitem{o3}
E.~H. Nam, B.~Kim, H.~Eom, and S.-L. Min.
\newblock Ozone ({O3}): An out-of-order flash memory controller architecture.
\newblock {\em TOC}, 2011.

\bibitem{exp:msr}
D.~Narayanan, E.~Thereska, A.~Donnelly, S.~Elnikety, and A.~Rowstron.
\newblock Migrating server storage to {SSDs}: analysis of tradeoffs.
\newblock In {\em EuroSys}, 2009.

\bibitem{int:onfi}
{ONFI Working Group}.
\newblock Open nand flash interface.
\newblock In {\em http://onfi.org/}.

\bibitem{park}
C.~Park, E.~Seo, J.-Y. Shin, S.~Maeng, and J.~Lee.
\newblock Exploiting internal parallelism of flash-based {SSD}s.
\newblock {\em IEEE CAL.}, 2010.

\bibitem{spec:pcie}
{PCI-SIG}.
\newblock {PCI} express base 3.0 specification.
\newblock 2012.

\bibitem{rafique}
N.~Rafique, W.-T. Lim, and M.~Thottethodi.
\newblock Effective management of {DRAM} bandwidth in multicore processors.
\newblock In {\em PACT}, 2007.

\bibitem{exp:iotta}
S.~Repository.
\newblock http://iotta.snia.org/.

\bibitem{int:sata}
{SATA-IO}.
\newblock {\em Serial ATA Revision 3.1}.
\newblock 2011.

\bibitem{spec:mavell}
{Shawn Kang}.
\newblock Native {PCI}e {SSD} controllers.
\newblock 2012.

\bibitem{shin}
J.-Y. Shin, Z.-L. Xia, N.-Y. Xu, R.~Gao, X.-F. Cai, S.~Maeng, and F.-H. Hsu.
\newblock {FTL} design exploration in reconfigurable high-performance {SSD} for
  server applications.
\newblock In {\em ICS}, 2009.

\bibitem{zuravleff}
W.~K. Zuravleff and T.~Robinson.
\newblock Controller for a synchronous {DRAM} that maximizes throughput by
  allowing memory requests and commands to be issued out of order.
\newblock {\em U.S. Patent No: 5,630,096}, 1997.

\end{thebibliography}

\end{document}